\documentclass[aps,prd,twocolumn,twoside,preprintnumbers,superscriptaddress,floatfix]{revtex4-2}

\usepackage{epsfig}
\usepackage{amsmath}
\usepackage{amssymb}
\usepackage{latexsym}
\usepackage{xspace}
\usepackage[colorlinks=true,linktocpage=true,linkcolor=blue,citecolor=blue,allcolors=blue]{hyperref}
\usepackage{url}
\usepackage[utf8]{inputenc}
\usepackage{indentfirst}
\usepackage{enumerate}
\usepackage{color}
\usepackage{relsize}
\usepackage{graphicx}
\usepackage{dcolumn}
\usepackage{bm}
\usepackage[caption=false,position=top]{subfig}
\usepackage[english]{babel}
\usepackage{slashed}

\newcommand{\be}{\begin{equation}}
\newcommand{\ee}{\end{equation}}
\newcommand{\bea}{\begin{eqnarray}}
\newcommand{\eea}{\end{eqnarray}}

\begin{document}

\title{Particle dispersion in the classical vector dark matter background}

\author{Eung Jin Chun}
 \email{ejchun@kias.re.kr}
 \affiliation{Korea Institute for Advanced Study, Seoul 02455,  South Korea}

 \author{Seokhoon Yun}
 \email{seokhoon.yun@pd.infn.it}
\affiliation{Dipartimento di Fisica e Astronomia, Universit\`a degli Studi di Padova, Via Marzolo 8, 35131 Padova, Italy}
\affiliation{INFN, Sezione di Padova, Via Marzolo 8, 35131 Padova, Italy}

\preprint{KIAS-P22022}
\date{\today}

\begin{abstract}

Interactions with a background medium modify in general the dispersion relation and  canonical normalization of propagating particles. This can have an important phenomenological consequence when considering light dark matter coupling to quarks and leptons.   In this paper, we address this issue in the vector dark matter background with the randomly distributed polarizations or a fixed polarization to the single direction. The observations associated with particle dispersion can give constraints on new light Abelian gauge boson models. Considering the solar neutrino transition and the electron mass measurement, stringent bounds can be put on  the gauged $L_\mu - L_\tau$ model and the dark photon model. Moreover, the classical vector field turns out to induce drastic changes in the particle  normalization, which rule out a significant parameter region of the generic vector dark matter model.

\end{abstract}

\maketitle


\section{Introduction}

Numerous phenomenological evidences for, e.g., the presence of dark matter (DM) advocate the point of view in particle physics that the Standard Model (SM) is not the end of the story.
A new $U(1)$ gauge symmetry~\cite{Fayet:1980ad,Fayet:1980rr,Holdom:1985ag,Fayet:1990wx} is typically accounted to be a minimal piece added to the SM, and it indeed appears ubiquitously in a number of well-motivated ultraviolet theories beyond the Standard Model.
A light and feebly interacting vector field of this new gauge symmetry, which we call a dark gauge boson dubbed as $\gamma^\prime$ or $A^\prime_\mu$, is of particular interest in phenomenological point of view. It can be a force mediator connecting  the visible sector with the hidden sector as the concept of a portal~\cite{Proceedings:2012ulb,Essig:2013lka,Raggi:2015yfk,Deliyergiyev:2015oxa,Alekhin:2015byh,Curciarello:2016jbz,Alexander:2016aln,Beacham:2019nyx,Fabbrichesi:2020wbt,Caputo:2021eaa}, and address some notable experimental anomalies such as the muon anomalous magnetic moment~\cite{Muong-2:2006rrc,Roberts:2010cj,Aoyama:2020ynm,Muong-2:2021ojo} and beryllium nuclear decays~\cite{Feng:2016jff,Feng:2016ysn}.
Although various experiments in the energy frontier are hard to detect these light and secluded vector bosons (and also scalars like axions), many ongoing and proposed tests in the intensity and cosmic Frontiers will achieve remarkable sensitivities to probe them in the near future.

Furthermore, such a light vector boson can be a viable cold DM candidate itself.
Several production mechanisms have been recently proposed, but the common feature is based on an oscillating homogeneous (vector-)bosonic field, which is a coherent state of particles and comprises a nonrelativistic component in the universe.
One approach to produce a classical vector field is the misalignment mechanism~\cite{Nelson:2011sf},
which is well known in the axion cold DM generation~\cite{Preskill:1982cy,Abbott:1982af,Dine:1982ah}.
As noted in Refs.~\cite{Arias:2012az,Graham:2015rva}, the large dark gauge boson coupling $R A^{\prime\mu} A^\prime_\mu /12$ to the scalar curvature is required for the misalignment to effectively generate the DM abundance.
Even without a nonminimal gravitational coupling, the vector dark matter (VDM) can be produced by quantum fluctuations during the inflationary phase due to the nonconformal coupling of the longitudinal polarization to gravity~\cite{Graham:2015rva}; the DM abundance depends on the dark gauge boson mass $(m_{\gamma^\prime})$ and the Hubble scale of inflation, then the current constraints on the inflationary scale~\cite{Planck:2018jri} demands $m_{\gamma^\prime} > 5\times 10^{-5}\,{\rm eV}$ in order to saturate the DM abundance.
The alternative way to produce the relic abundance of vector DM over a wide range of its mass is the tachyonic production mediated by an initially oscillating axion (or axion-like particle) field~\cite{Agrawal:2018vin,Dror:2018pdh,Co:2018lka,Bastero-Gil:2018uel,Co:2021rhi}.
When the axion field starts oscillating, its anomalous coupling to the dark gauge boson induces a tachyonic instability in the equation of motion of $A^\prime_\mu$ that dissipates and transfers the energy density of the initial axion condensate into the dark gauge boson with a specific helicity.
Moreover, the decay of topological defects (e.g., a  network of cosmic strings~\cite{Long:2019lwl}) can contribute to the VDM abundance.

In the presence of a background medium, the dispersion relation as well as normalization of a propagating particle can be modified ~\cite{Silin:1960pya,1962NCim...25..428J,Beaudet:1967zz,Klimov:1982bv,Weldon:1982aq,Braaten:1993jw}.
One familiar case is photons in an ionized plasma (e.g., the cosmological thermal bath or the circumstance inside astrophysical compact objects) or a material (e.g., water) that experience an index of refraction, which deviates from unity due to their coherent interactions with charged particles in the medium.
In the effective theory framework, quantized electromagnetic excitations in the medium are interpreted as the superposition of all scattered waves, and could acquire an effective mass in the dispersion which is a a complicated function of momentum due to the Lorentz symmetry breaking by singling out an inertial frame.
Likewise, when a particle propagates in a bosonic field background, its coupling to such a boson could alter its dispersion relation.
We note that the effective mass of fermions in the dispersion is distinguished from the (chirality breaking) bare mass which can be generated, e.g., by the vacuum expectation value of the Higgs field in SM.
For instance, neutrinos propagating in a scalar field background can obtain effective mass-squared which is chirality preserving~\cite{Choi:2019zxy,Choi:2020ydp}.

\medskip

In this paper, we discuss the medium effect on the particles propagating in a (ultra-)light VDM background.
 Barring the model-dependence in terms of the DM origin, we focus on the late time cosmology (i.e., $H\ll m_{\gamma^\prime}$) where bosons already roll down and start to oscillate around the minimum of the potential.
Contrary to a scalar or axion DM,  there is an issue of the polarization.
In some of the scenarios discussed above (e.g., the production via the misalignment mechanism or a tachyonic instability), a specific direction is imposed on DM within the cosmological horizon, and a degree of such a preferred polarization may remain unchanged for most of the universe history~\cite{Caputo:2021eaa,Arias:2012az}.
However, the effect of cosmological structure formation on the DM polarization is unclear so far~\footnote{Recently, the impact of the small scale structure in vector dark matter on its polarizations has been discussed in Ref.~\cite{Amin:2022pzv}.}, and the randomized polarization can also be accomplished in some scenarios (e.g., the production from cosmic strings that the decay of long strings and the collapse of short loops could distribute the DM polarizations in a democratic way).
Hence, we adopt a phenomenological approach as in Ref.~\cite{Caputo:2021eaa} that the two extreme cases are taken into account; the VDM with a fixed single polarization (``{\it polarized background}") or the equally distributed polarizations (``{\it unpolarized background}").
As we will see, the behavior of the modified dispersion relation is similar in both cases.
Furthermore, we are interested in the observations without directional information that accounts for the robustness of our results. For our discussion, we will consider the popular examples of anomaly-free Abelian symmetries such as kinetically mixed $U(1)$, $B-L$, and $L_\mu-L_\tau$ and so on.

The paper is organized as follows.
Section~\ref{sec:Dispersion} is devoted to the analysis of the particle dispersion and normalization in the classical VDM for the two limited scenarios: the unpolarized and polarized background.
In Sec.~\ref{sec:Cons}, we confront our findings with the phenomenological observations associated with such a refractive phenomenon and derive constraints on some gauged $U(1)$ extensions of SM.
We then provide discussions and our conclusion in Sec.~\ref{sec:conclusions}.

\section{Dispersion of particles in the medium of vector dark matter}
\label{sec:Dispersion}

In this section, we discuss the particle dispersions in the VDM background.
Renormalizable interactions to describe the interplay of the dark gauge boson $\gamma^\prime$ and a SM fermion $\psi$ are written as
\bea
\mathcal{L}_{\gamma^\prime \text{-}\psi} =  g_\psi A^\prime_\mu  \bar{\psi} \Gamma^\mu \psi \,
\eea
with the interaction strength $g_\psi$.
The matrix $\Gamma^\mu$ becomes $\gamma^\mu$ and $\gamma^\mu\gamma^5$ for the vector and axial-vector current coupling cases, respectively, but can be in general a linear combination of them.
In this paper, we investigate the vector and axial-vector current coupling cases, the results of which can be easily applied to the generic cases.

Furthermore, even if the SM sector is not explicitly charged under the dark $U(1)$ gauge symmetry, the gauge boson of the dark $U(1)$ can still interact with the SM particles via the kinetic mixing~\cite{Holdom:1985ag} (the so-called vector portal), which is allowed for Abelian gauge symmetries.
The Lagrangian to describe such kinetic mixing reads
\be
\mathcal{L}_{ \gamma^\prime\text{-}\gamma} = \frac{\varepsilon}{2}F_{\mu\nu}F^{\prime \mu\nu} \, ,
\label{eq:KineticMix}
\ee
where $F_{\mu\nu} = \partial_\mu A_\nu - \partial_\nu A_\mu$ and $F_{\mu\nu}^{\prime} = \partial_\mu A^\prime_\nu - \partial_\nu A^\prime_\mu$ are the field strength of the SM photon $A^\mu$ and the dark gauge boson $A^{\prime\mu}$, respectively, and $\varepsilon$ the dimensionless kinetic mixing parameter.
The effective couplings to the SM particles through the kinetic mixing follow the form of the vector current with the electromagnetic charge, i.e. $\Gamma^\mu \rightarrow \gamma^\mu$ and $g_\psi \propto e q_\psi$ with $e$ (minus) the electron charge and $q_\psi$ the electric charge of $\psi$.
Therefore, the dark gauge boson in the presence of the only kinetic mixing is dubbed the dark photon~\cite{Arkani-Hamed:2008hhe}.

\begin{figure}[h!]
\centering
\includegraphics[scale=.65]{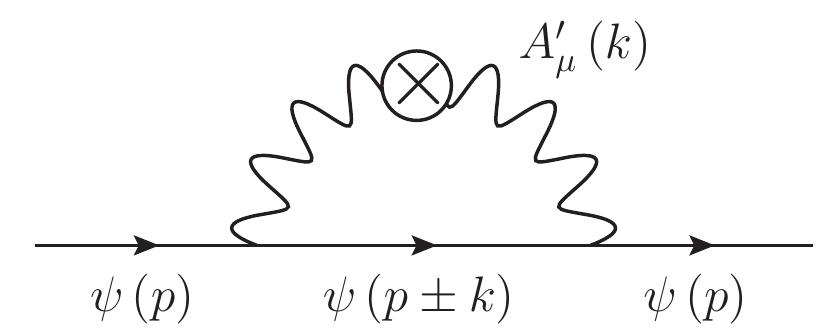}
\caption{Self-energy diagram of a $\psi$ field (solid) in the vector dark matter background (wavy with $\otimes$). The momentums of the external $\psi$ legs and the dark matter are denoted by $p$ and $k$, respectively. The momentum of the intermediate $\psi$ line $p\pm k$ accounts for the contribution from the dark matter or antidark matter distribution.}
\label{fig:SelfE}
\end{figure}

The leading order self-energy diagram to account for the medium effect on the effective 2-point function of $\psi$ in the VDM background is shown in Fig.~\ref{fig:SelfE}.

The solid lines and the wavy line represent  the $\psi$ and dark gauge boson field, respectively.
We mark $\otimes$ on the wavy line to indicate the propagator of the VDM in terms of the  classical field~\footnote{There is also the correction to the 2-point function from the same diagram of Fig.~\ref{fig:SelfE} with the quantum loop (i.e., the quantum flunctuated propagator with a loop momentum). Due to the loop suppression, we ignore this radiatively induced contribution.}.
The four momentum of the external $\psi$ lines and the (on-shell) VDM is denoted by $p$ and $k$, respectively; when the momentum $k$ points into the left (right) vertex in Fig.~\ref{fig:SelfE}, the intermediate $\psi$ line has $p+k$ ($p-k$) from the momentum conservation.
The self-energy diagram in Fig.~\ref{fig:SelfE} results in the correction to the 2-point operator, which reads as follows
\bea
\mathcal{L}_\psi^{\rm eff} = \bar{\psi}i\slashed{\partial}\psi - m_\psi \bar{\psi}\psi - \bar{\psi}\slashed{\Sigma}\psi \,
\label{eq:EffLag}
\eea
with the amplitude
\bea
-i \slashed{\Sigma} & = & \left(-i g_f^2\right)\int \frac{d^4 k}{\left(2\pi\right)^4} \Gamma^\mu \frac{\left(\slashed{p}+\slashed{k} + m_\psi\right)}{\left(p+k\right)^2-m_\psi^2} \Gamma^\nu \Delta_{\mu\nu}^{\gamma^\prime} \, ,
\label{eq:Amp}
\eea
where $m_{\psi}$ is the bare $\psi$ mass, and $\Delta_{\mu\nu}^{\gamma^\prime}$ accounts for the propagator of the dark gauge boson in the loop.
The background part of $\Delta_{\mu\nu}^{\gamma^\prime}$ can be written as follow
\bea
\left.  \Delta_{\mu\nu}^{\gamma^\prime}\right|_{\rm DM} = \sum_{a} \epsilon^{a*}_\mu \epsilon^a_\nu \Big(2\pi \delta(k^2-m_{\gamma^\prime}^2) f_{\gamma^\prime}^a\Big) \, ,
\label{eq:DMprop}
\eea
where and $f_{\gamma^\prime}^a$ is the DM distribution function of a polarization state $a$.
The term in the parenthesis of Eq.~\eqref{eq:DMprop} can be interpreted as the spectral density function of the DM (for one-particle states).
We assume the distribution function to be monochromatic
\bea
f_{\gamma^\prime}^a = \left(2\pi\right)^3 \delta^{(3)}[\vec{k}-\vec{k}_{\gamma^\prime}] \Big(\Theta (k_0)n_{ \gamma^\prime} + \Theta (-k_0)n_{\gamma^\prime} \Big) \xi^a \, , \quad
\label{eq:dpdmDis}
\eea
where $\vec{k}_{\gamma^\prime}$ is the spatial momentum of the VDM background, $n_{\gamma^\prime}$ the total DM number density with the fraction $\xi^a$ for the polarization $a$ ($\sum_a\xi^a = 1$).
Here the real condition of vector bosons imposes $n_{\gamma^\prime} = n_{\bar{\gamma}^\prime}$ where $\bar{\gamma}^\prime$ is the anti-particle of $\gamma^\prime$.
Since the DM is nonrelativistic as $m_{\gamma^\prime} \gg |\vec{k}_{\gamma^\prime}|$, the spatial momentum of the DM gives a negligible effect on the particle dispersions that the results would not hinge on the explicit distribution function.
Therefore, our assumption in Eq.~\eqref{eq:dpdmDis} turns out to be plausible.
We dub $k_{\gamma^\prime}$ the monochromatic four momentum of the VDM background in order to distinguish it from $k$ the integration four momentum in Eq.~\eqref{eq:Amp}.

\subsection{Unpolarized background}
\label{sec:unpol}

When the DM distribution is independent of polarizations as $\xi^a = 1/3$ for all $a$, the polarization sum can be prescribed as the metric tensor $\sum \epsilon^*_\mu \epsilon_\nu \rightarrow -g_{\mu\nu} + k_\mu k_\nu / m_{\gamma^\prime}^2$.
The classical vector field background breaks the gauge symmetry, hence the amplitude of the self-energy diagram in Fig.~\ref{fig:SelfE} does not follow the Ward-Takahashi identity~\cite{Ward:1950xp,Takahashi:1957xn} that the term of $k_\mu k_\nu / m_{\gamma^\prime}^2$ in the polarization sum gives the finite contribution.
The amplitude is given by
\bea
-i \slashed{\Sigma} & \simeq & -i\left(\slashed{p} \Sigma_p + \slashed{k}_{\gamma^\prime} \Sigma_k \mp m_{\psi} \Sigma_m \right)  \,
\eea
with $-$ ($+$) sign in the last term for the vector (axial-vector) current and
\bea
\Sigma_p & = & \frac{\delta m_\psi^2}{3}\frac{\Delta+m_{\gamma^\prime}^2}{\left(\Delta+m_{\gamma^\prime}^2\right)^2-4m_{\gamma^\prime}^2E^2}\, ,\\
\Sigma_k & = & \frac{2\delta m_\psi^2}{3}\frac{\Delta-2m_{\gamma^\prime}^2}{\left(\Delta+m_{\gamma^\prime}^2\right)^2-4m_{\gamma^\prime}^2E^2} \left(\frac{E}{m_{\gamma^\prime}}\right)\, , \\
\Sigma_m & = & 3 \Sigma_p \, ,
\eea
where we define
\be
\Delta=E^2-|\vec{p}|^2 - m_\psi^2
\label{eq:Delta}
\ee
with $p^\mu = \left(E,\vec{p}\right)$.
The $\delta m_\psi^2$ factor parametrizes how coherent interactions with the DM contribute to modifications of the dispersion, and reads as follows
\bea
\delta m_\psi^2 =  g_\psi^2 \frac{\rho_{\gamma^\prime}}{m_{\gamma^\prime}^2}
\label{eq:delta}
\eea
with $\rho_{\gamma^\prime} = m_{\gamma^\prime} n_{\gamma^\prime}$ the energy density of the VDM and $m_{\gamma^\prime}$ the dark gauge boson mass.

The effective Lagrangian in Eq.~\eqref{eq:EffLag} gives the equation of motion for $\psi$ as follows
\bea
\slashed{p}-m_\psi - \slashed{\Sigma} & \simeq & \gamma^0 \Big(E\left(1-\Sigma_p\right) - m_{\gamma^\prime}\Sigma_k\Big) - \vec{\gamma}\cdot\vec{p} \Big(1-\Sigma_p\Big) \nonumber\\
&& - m_\psi\Big(1\mp \Sigma_m\Big) = 0 \, .
\eea
In order to account for the dispersion relation, which is parametrized by $\Delta$ in Eq.~\eqref{eq:Delta}, we need to solve
\be
\begin{split} \label{eq:disp}
& \Big(E \left(1-\Sigma_p\right)- m_{\gamma^\prime}\Sigma_k\Big)^2 \\
 = & \left|\vec{p}\right|^2 \Big(1-\Sigma_p\Big)^2 + m_\psi^2\Big( 1- \alpha_m\Sigma_p\Big)^2\, ,
 \end{split}
\ee
where $\alpha_m = \pm 3$ for the vector and axial-vector current coupling cases, respectively.
We find the full expression in expansion with respect to $\Delta$ as follows
\bea
\sum_{i=0}^5 \Upsilon_i \Delta^i = 0 \,
\label{eq:DeltaEq}
\eea
with
\bea
\Upsilon_0 & = & \frac{1}{9}m_{\gamma^\prime}^4 \delta m_\psi^2 \Bigg[ \delta m_\psi^2 \Big(8|\vec{p}|^2+\left(9-\alpha_m^2\right)  m_\psi^2\Big) \label{eq:Upsilon0}\\
&& +6 \Big(m_{\gamma^\prime}^2-4 m_\psi^2 - 4|\vec{p}|^2\Big) \Big(4 |\vec{p}|^2 + \left(3+ \alpha_m\right)m_\psi^2\Big) \Bigg] \, ,\nonumber \\
\Upsilon_1 &=& m_{\gamma^\prime}^2 \Bigg[-\frac{2}{9}\delta m_\psi^2 \Big(\delta m_\psi^2 \left(10 |\vec{p}|^2+\left(9+\alpha_m^2\right) m_\psi^2 \right) \nonumber\\
&& -12\left(m_\psi^2+|\vec{p}|^2\right)\left(2|\vec{p}|^2+\left(3-\alpha_m\right) m_\psi^2\right)\Big) \nonumber   \\
&& +\frac{1}{3}m_{\gamma^\prime}^2\Big(48\left(m_\psi^2+|\vec{p}|^2\right)^2 \nonumber \\
&&-2\delta m_\psi^2 \left(22 |\vec{p}|^2 + \left(21+\alpha_m\right)m_\psi^2\right) + 3\delta m_\psi^4\Big) \nonumber \\
&&  + 2 m_{\gamma^\prime}^4 \Big(-4\left(m_\psi^2 + |\vec{p}|^2\right)+\delta m_\psi^2\Big)+m_{\gamma^\prime}^6\Bigg] \, , \\
\Upsilon_2 &=&  \frac{8}{9}\delta m_\psi^4 \Big( |\vec{p}|^2 + \frac{9-\alpha_m^2}{8} m_\psi^2\Big)  \nonumber \\
&& - \delta m_\psi^2 m_{\gamma^\prime}^2 \Big(2\delta m_\psi^2-\frac{42-2\alpha_m}{3} m_\psi^2 - \frac{40}{3} |\vec{p}|^2\Big)  \ \nonumber \\
&& + 2m_{\gamma^\prime}^4 \Big(8 \left(m_\psi^2+|\vec{p}|^2\right)-3 \delta m_\psi^2\Big) -4 m_{\gamma^\prime}^6 \, ,\\
\Upsilon_3 &= &  \delta m_\psi^4 - \frac{2}{3}\delta m_\psi^2\Big( 2 |\vec{p}|^2 + \left(3- \alpha_m\right)  m_\psi^2\Big) \nonumber \\
&& +m_{\gamma^\prime}^2 \Big(6\delta m^2-8\left(m_\psi^2 +  |\vec{p}|^2\right)\Big) +6 m_{\gamma^\prime}^4   \, , \\
\Upsilon_4 &=& -2 \delta m_\psi^2 -4 m_{\gamma^\prime}^2  \, , \qquad \Upsilon_5 = 1 \, . \label{eq:Upsilon5}
\eea
Since the equation in Eq.~\eqref{eq:DeltaEq} is quintic with the coefficients given in Eqs.~\eqref{eq:Upsilon0}-\eqref{eq:Upsilon5}, there are basically five solutions of $\Delta$.
Note that the unique branch of the solutions fulfills the relevant conditions; the solution is real and positive, and vanishes when $g_\psi\rightarrow 0$.
We describe the value of $\Delta$ in this branch for some limits, which cover the cases of interest.
The DM background singles out an inertial frame to break Lorentz invariance (also gauge invariance as discussed previously) that accounts for the effective dispersion relations in function of the momentum.

\medskip

We first focus on the case of phenomenological interest; $m_\psi > \delta m_\psi, m_{\gamma^\prime}$ that corresponds to the realistic parameter space of the condensed VDM.  It will be interesting to see that the SM fermions develop distinctive dispersion relations depending on whether they have vector or axial-vector coupling to the DM.
The full momentum dependence of the dispersion relations, $\Delta(|\vec{p}|)$, can be obtained numerically for given values of the parameters.
We will show some sample plots later. Let us now present approximate analytic expressions valid at particular limits.

\underline{\it Vector coupling}:

Two regimes of the parameter space distinguished by the hierarchy between $\delta m_\psi^2$ and $m_\psi m_{\gamma^\prime}$ show different behaviors of the dispersion relations.
In the regime of $\delta m_\psi^2 < m_\psi m_{\gamma^\prime}$, we find
\begin{equation} \label{Vsol2}
\Delta \approx \left\{
\begin{split}
& \frac{2  \delta m_\psi^2 m_{\gamma^\prime}^2}{\delta m_\psi^2 + 3 m_{\gamma^\prime}^2} & \mbox{for}~~ |\vec{p}|\gg  m_\psi \\
& \delta m_\psi^2 & \mbox{for}~~ |\vec{p}|\ll  m_\psi
\end{split}
\right.\,.
\end{equation}
In the opposite regime of $\delta m_\psi^2 > m_\psi m_{\gamma^\prime}$, we get
\begin{equation} \label{Vsol1}
\Delta \approx \left\{
\begin{split}
& {2\over \sqrt{3}} \delta m_\psi |\vec{p}| & \mbox{for}~~ |\vec{p}|\gg \delta m_\psi \\
& \delta m_\psi^2 & \mbox{for}~~ |\vec{p}|\ll \delta m_\psi
\end{split}
\right.
\end{equation}
which has a totally different behavior at the relativistic limit.
\smallskip

\underline{\it Axial-vector  coupling}:

It has a peculiar behavior that a common solution exists for the both regimes:
\begin{equation} \label{Asol2}
\Delta \approx \left\{
\begin{split}
& \frac{2  \delta m_\psi^2 m_{\gamma^\prime}^2}{\delta m_\psi^2 + 3 m_{\gamma^\prime}^2} & \mbox{for}~~ |\vec{p}|\gg  m_\psi \\
& {2\over3} \frac{\delta m_\psi^2 m_{\gamma^\prime}^2}{\delta m_\psi^2 + m_{\gamma^\prime}^2}
\frac{|\vec{p}|^2}{m_\psi^2}  & \mbox{for}~~ |\vec{p}|\ll  m_\psi
\end{split}
\right. \,,
\end{equation}
and there appears an additional solution
in the regime of  $\delta m_\psi^2 >  m_{\gamma^\prime}^2$:
\begin{equation} \label{Asol1}
\Delta \approx \left\{
\begin{split}
& {2\over \sqrt{3}} \delta m_\psi |\vec{p}| & \mbox{for}~~ |\vec{p}|\gg  m_\psi \\
& 2 \delta m_\psi m_\psi  + \delta m^2_\psi  & \mbox{for}~~ |\vec{p}|\ll  m_\psi
\end{split}
\right. \, .
\end{equation}
Notice that the dispersion relations at the relativistic limit are the same for the vector and axial-vector couplings.

For the sake of completeness, let us consider the strong background effect allowing $\delta m_\psi^2 >  m_\psi^2$ which is not of phenomenological interest.
For the vector (axial-vector) coupling of DM, the solution of $\Delta$ is found to be
\begin{equation} \label{eq:strong}
\Delta \approx \left\{
\begin{split}
& {2\over \sqrt{3}} \delta m_\psi |\vec{p}|    &  \mbox{for}~~ |\vec{p}|\gg \delta m_\psi \\
& \delta m_\psi^2  \left( +2 \delta m_\psi  m_\psi \right) & \mbox{for}~~ |\vec{p}|\ll   \delta m_\psi
\end{split}
\right.
\end{equation}
which is basically the same as the solution \eqref{Vsol1} or \eqref{Asol1}.
We remark that there appears a crucial difference between dispersion relations in the vector and scalar background. In the ultrarelativistic limit, the dispersive behavior of fermions in the vector medium  shows  $\Delta \propto |\vec{p}|$ as discussed above, while it goes like $\Delta \sim \delta m^2_\psi$ in the scalar medium \cite{Choi:2019zxy,Choi:2020ydp}.
\medskip

Another important medium effect is the modification of field normalization of the particles interacting with the medium. Generalizing the discussion of  \cite{Weldon:1982aq}, we find the normalization factor $Z$ given by
\begin{equation}
Z = \left[ {\partial \over \partial E} (V_0 - \sqrt{V_p^2 + M^2})\right]^{-1}_{\mbox{pole}} \, ,
\end{equation}
where $V_0 \equiv E(1-\Sigma_p)-m_{\gamma'} \Sigma_k$, $V_p= |\vec{p}|(1-\Sigma_p)$, and $M\equiv m_\psi(1\mp\Sigma_m)$ considering the positive energy solution of~\eqref{eq:disp}.
The fermions following the dispersion relations given in Eqs.~\eqref{Vsol2}-\eqref{eq:strong} have also distinct normalization factors.
\smallskip

\underline{\it Vector coupling}:

The behavior of the normalization factor for the solution (\ref{Vsol2}) is  distinguished further by the hierarchy between  $\delta m_\psi^2$ and $m_{\gamma^\prime}^2$. In the regime of $\delta m_\psi^2\ll m_{\gamma^\prime}^2 < m_\psi m_{\gamma^\prime}$, the normalization factor becomes trivial ($Z\to 1$) manifesting the suppressed medium effect.
In the opposite regime of $m_{\gamma^\prime}^2 < \delta m_\psi^2 < m_\psi m_{\gamma^\prime}$, the normalization factor becomes momentum-dependent and shows a different limiting behavior:
\begin{equation} \label{VZ2}
Z \approx \left\{
\begin{split}
& \frac{ 3m_{\gamma^\prime}^2}{\delta m_\psi^2+3m_{\gamma^\prime}^2}  & ~\mbox{for}~~ |\vec{p}|\gg  m_\psi \\
& ~ 1 & ~ \mbox{for}~~ |\vec{p}| \ll m_\psi
\end{split}
\right. \, .
\end{equation}
The explicit transition behavior will be shown later in the sample plots together with $\Delta$.
For the solution (\ref{Vsol1}) applicable to the regime $\delta m_\psi^2 >  m_\psi m_{\gamma^\prime}$, we have
\be \label{VZ1}
Z \approx \left\{
\begin{split}
& 1/2   & \mbox{for}~~|\vec{p}|\gg \delta m_\psi \\
&    \frac{ \delta m_\psi^2/2}{2 m_\psi^2+ \delta m_\psi^2}  &  \mbox{for}~~
|\vec{p}| \ll \delta m_\psi
\end{split}
\right. \, .
\ee
\smallskip

\underline{\it Axial-vector coupling}:

It is interesting to see that the normalization factors are almost momentum-independent in this case.
Depending on the hierarchy between $\delta m_\psi^2$ and $m_{\gamma'}^2$, the normalization factor has two different values corresponding to the dispersion relations \eqref{Asol2} and \eqref{Asol1}, respectively:
\bea
Z & \approx &  1   \qquad \quad  \mbox{for}~~ \delta m_\psi^2 <  m_{\gamma'}^2 \, ,  \label{AZ2}  \\
Z & \approx &  1/2  \qquad   \mbox{for}~~ m_{\gamma'}^2 < \delta m_\psi^2  \label{AZ1}
 \, .
\eea

A notable feature is that we always have $Z\approx1$ for  $\delta m_\psi \ll  m_{\gamma'}$, and $Z\approx1/2$ for the branches with $\Delta \propto \delta m_\psi |\vec{p}|$.  Furthermore,
the strong background of $\delta m_\psi^2 \gg  m_\psi^2$ (\ref{eq:strong}) allows the constant normalization  factor $Z\approx 1/2$ for both the vector and axial-vector coupling.
\smallskip

\begin{figure}[h!]
\centering
\includegraphics[scale=.65]{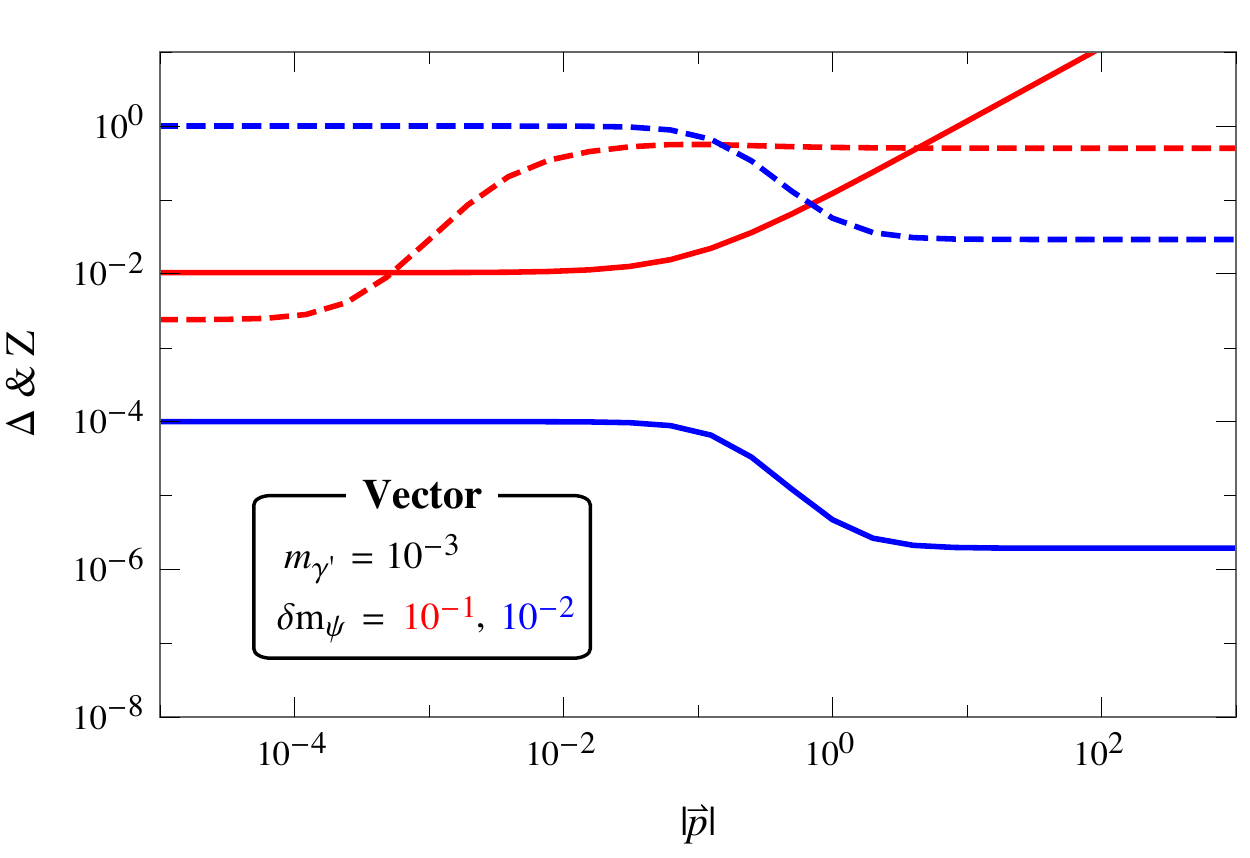}
\includegraphics[scale=.65]{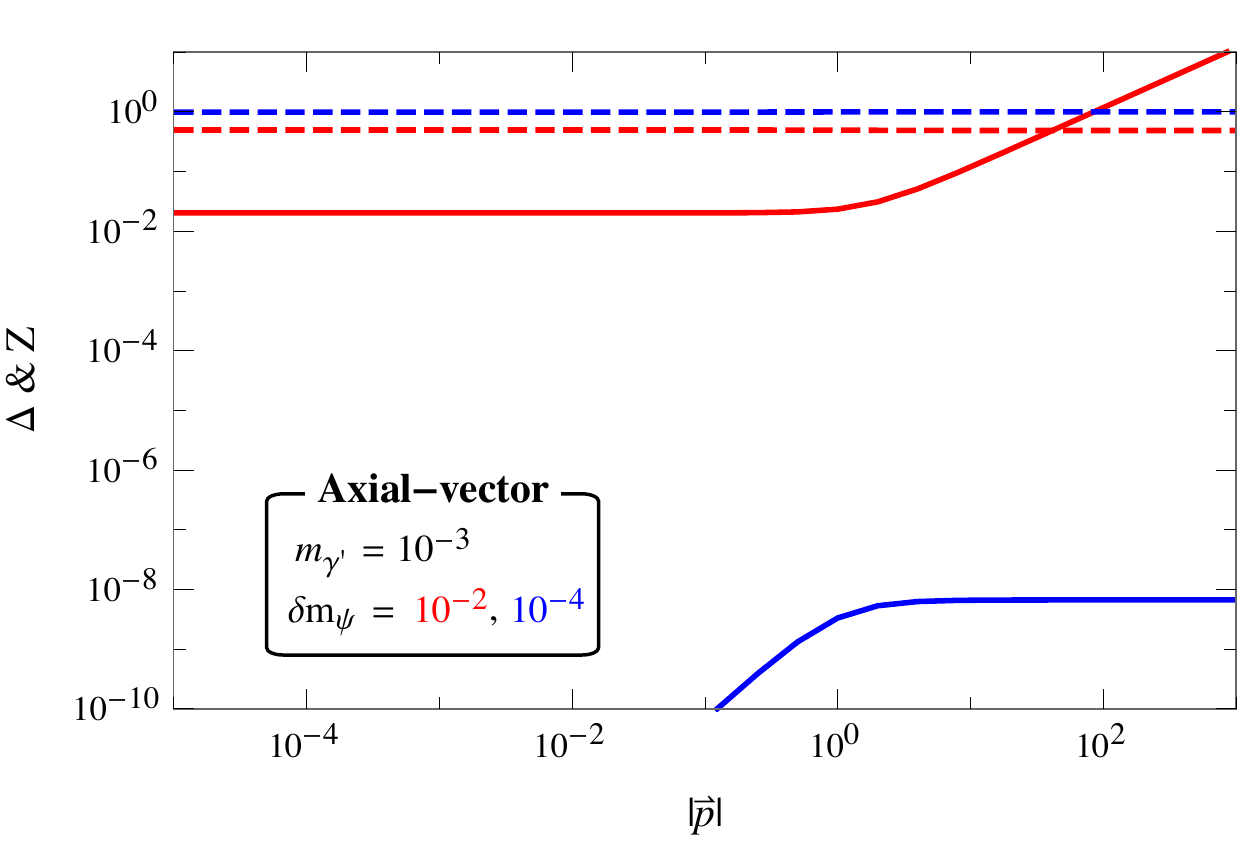}
\caption{Examples of numerical solution of the modification to the particle dispersion relation $\Delta$ and the normalization factor $Z$.  All the mass parameters are in unit of $m_\psi$.  In the upper panel, the blue (red) solid and dashed lines correspond to $\Delta$ in \eqref{Vsol2} [\eqref{Vsol1}] and $Z$ in \eqref{VZ2} [\eqref{VZ1}], respectively.
In the lower panel,  the blue (red) solid and dashed lines correspond to $\Delta$ in \eqref{Asol2} [\eqref{Asol1}] and $Z$ in \eqref{AZ2} [\eqref{AZ1}], respectively.}
\label{fig:VA-DZ}
\end{figure}

Let us now present the plots for the full momentum dependence of $\Delta$ and $Z$ to see the transition behaviors between the limiting solutions discussed above. Figure~\ref{fig:VA-DZ} shows the numerical solutions obtained with $m_{\gamma'}/m_\psi=10^{-3}$ and  two sample values of $\delta m_\psi/ m_\psi$; $10^{-1}\, (\delta m^2_\psi > m_\psi m_{\gamma'})$ and $10^{-2}\, (\delta m^2_\psi < m_\psi m_{\gamma'})$ for the vector coupling in the upper panel, and $10^{-2}\, (\delta m_\psi^2 > m_{\gamma'}^2)$ and $10^{-4}\, (\delta m_\psi^2 < m_{\gamma'}^2)$ for the axial-vector coupling in the lower panel.
The former  corresponds to the red lines, and the latter to the blue lines.  The solid and dashed lines show the momentum dependence of $\Delta/m_\psi^2$ and $Z$, respectively.

\subsection{Polarized background}
\label{sec:pol}

In the assumption that the classical VDM background prefers a specific polarization, we set such a polarization as $z$-direction without loss of generality so $\xi^3 = 1$.
We consider the homogeneous background at the leading order that $\partial_i A^\mu = 0$ implies $A^0 = 0$.
Following the same steps in the previous subsection, the amplitude of the self-energy diagram is written by
\bea
-i \slashed{\Sigma} & \simeq & -i\left[\left(\slashed{p} +2|\vec{p}_z| \gamma^3\right)\tilde{\Sigma}_p + \slashed{k}_{\gamma^\prime} \tilde{\Sigma}_k \mp m_{\psi} \tilde{\Sigma}_p \right]  \,
\eea
with
\bea
\tilde{\Sigma}_p & = &  \delta m_\psi^2\frac{\Delta+m_{\gamma^\prime}^2}{\left(\Delta+m_{\gamma^\prime}^2\right)^2-4m_{\gamma^\prime}^2E^2} \, ,\\
\tilde{\Sigma}_k & = & \delta m_\psi^2\frac{-2m_{\gamma^\prime}E}{\left(\Delta+m_{\gamma^\prime}^2\right)^2-4m_{\gamma^\prime}^2E^2} \, ,
\eea
and $\vec{p}_i =(\hat{i}\cdot \vec{p})\, \hat{i} $.
The equation of motion with the self-energy term induces the dispersion relation as following
\be
\begin{split}
& \Bigg(E \left(1-\tilde{\Sigma}_p\right)- m_{\gamma^\prime}\tilde{\Sigma}_k\Bigg)^2 - m_\psi^2\left( 1- \tilde{\alpha}_m\tilde{\Sigma}_p\right)^2 \\
 = & \sum_{i\neq z}\left|\vec{p}_i\right|^2 \Big(1-\tilde{\Sigma}_p\Big)^2  + \left|\vec{p}_z\right|^2 \Big(1+\tilde{\Sigma}_p\Big)^2 \, ,
\end{split}
\label{eq:AXEoM}
\ee
where $\tilde{\alpha}_m = \pm 1$ for the vector and axial-vector current coupling cases, respectively.

 Finding the proper solution for $\Delta$ are summarized in Appendix.
Except the extreme case of $\delta m_\psi^2 \cos^2\theta \ll  m_{\gamma^\prime}^2$ with $\cos\theta = \hat{z}\cdot \vec{p}/|\vec{p}|$, we appreciate that the modification of the dispersion relation is similar to the unpolarized case in each kinematic region.
Indeed, averaging over the angular information, which replaces $\cos^2\theta$ with $1/3$, gives the consistent results with the unpolarized background.
At the next section, we evaluate the constraints on the classical VDM from the phenomenological observations where such a directional information of a momentum is unimportant.
In this context, we take into account the dispersion relation in the unpolarized VDM in the remainder of this paper.

\section{Constraints}
\label{sec:Cons}

Let us discuss the constraints on the classical VDM from the phenomenologies associated with the particle dispersion and normalization.
The scenarios with anomaly-free gauged $U(1)$ gauge extensions, which are ubiquitous in UV models and appealing for some phenomenological aspects, are examined: the lepton flavor-dependent symmetries such as $U(1)_{L_e - L_{\mu}}$, $U(1)_{L_e-L_\tau}$, and $U(1)_{L_\mu - L_\tau}$, the flavor-universal  $U(1)_{B-L}$ (it is anomaly free if right-handed neutrinos are introduced).
We also investigate the so-called dark photon model where the kinetic mixing of an additional gauge boson with the SM photon is the only source for its interaction to the SM particles.

There are a few important features of the medium effect providing strong constraints on the model parameter space.
\begin{itemize}
\item In the cases of the vector coupling or the axial-vector coupling with $\delta m_\psi^2 > m_\psi m_{\gamma^\prime}$, the fermion $\psi$ develops an medium induced mass-squared $\Delta=\delta m^2_\psi$ or $\delta m_\psi m_\psi$ at the small momentum limit $\delta m_\psi^2 \gg |\vec{p}|^2$. This may contract with  the measurements of the fermion mass $m_\psi$ described in the SM.
\item For $\delta m_\psi^2 \gg m_\psi m_{\gamma^\prime}$ with $|\vec{p}|^2 \gg \delta m_\psi^2$, the DM contribution to the particle dispersion is quantified by $\Delta =\sqrt{4/3}|\vec{p}|\delta m_\psi$ as shown in Eq.~\eqref{Vsol1}. Thus, it gives a constant shift to the energy $\delta E_\psi \sim \delta m_\psi$ which may alter neutrino oscillations significantly if flavor-dependent.
\item If the normalization factor for the fermions interacting with VDM deviates from the trivial value $Z=1$, it may lead to an observable field/flavor dependence of the SM couplings which has been tested precisely.
\end{itemize}

As the first example, we can employ the precise measurement of the electron mass in the ground experiments and the relativistic degree of neutrinos at the epoch of matter-radiation equality $z_{\rm eq} \simeq 3400$.
In most of the experimental setups to measure the electron mass, electrons are well electromagnetically trapped inside a system (e.g., penning traps~\cite{PENNING1936873,PhysRev.76.565.2}) that their velocity is less than $10^{-3}c$~\cite{Brown:1985rh}.
Moreover, since the distribution of the DM in the Earth reference frame translates into an characteristic velocity of $\mathcal{O}(10^{-3})c$~\cite{Kerr:1986hz,Reid:2009nj,McMillan:2009yr,Bovy:2009dr,Freese:2012xd}, one can consider that electrons in the experiments are nonrelativistic with respect to the VDM.
The well-measured electron  mass ($m_{e,\rm meas} \simeq 0.511 {\rm MeV}$~\cite{Zyla:2020zbs}) is to be accounted by $m_{e,\rm meas}=\sqrt{m_e^2 +\delta m_e^2}$  where the bare and medium-induced mass-squared  are positive-definite by construction.
As a consequence, one can apply the conservative constraint: $ \delta m_e^2 < m_{e,\rm meas}^2$ that leads to
\be
g_e < 3.3 \times 10^{-12}  \Bigg(\frac{m_{\gamma^\prime}}{10^{-20}\,{\rm eV}}\Bigg) \Bigg(\frac{0.3\,{\rm GeV}/{\rm cm}^{3}}{\rho^\oplus_{\rm DM}}\Bigg)^{1/2}  \, , 
\label{eq:BoundEmass}
\ee
where $\rho^\oplus_{\rm DM}$ indicates the local DM density with $0.3\,{\rm GeV}/{\rm cm}^3$ at the $2\text{-}\sigma$ level~\cite{Benito:2019ngh}.
Note that this constraint can be further improved if the electron dispersion relation is evaluated in distinct environments with respect to the dark matter background.
As an example, since the deviation of the dispersion relation is momentum-dependent as discussed in the previous section, a robust bound may arise from a comparison between electron mass estimates at different velocities with respect to the DM background.

In the cosmology side, neutrinos also get an effective mass contribution from the DM, and $\delta m_\nu^2$ with respect to the bath temperature $T$ reads as follows
\bea
\delta m_\nu^2 \left(T\right) & = & g^2_\nu \frac{\rho_{\rm DM}^0}{m_{\gamma^\prime}^2}  \left(\frac{g_{*s}(T) T^3}{g_{*s}(T_0) T_0^3}\right)   \, ,
\eea
where $\rho_{\rm DM}^0 = 1.26 \,{\rm keV}/{\rm cm}^{3}$ denotes the current DM density, $g_{*s}(T)$ the entropic degrees of freedom, and $T_0 =2.726\,{\rm K}$~\cite{Mather:1998gm,Fixsen:2009ug,Noterdaeme:2010tm} the current cosmic microwave background (CMB) temperature.
In order for relic neutrinos to compose the background radiation at the matter-radiation equality ($T_{\rm eq}=z_{\rm eq}T_0 \simeq 0.80\,{\rm eV}$) appropriately, as a naive estimation, $\delta m_\nu$ should not exceed the temperature of neutrinos $T_\nu = (4/11)^{1/3} T$, then we obtain the constraint
\bea
g_\nu & < &  9.3\times 10^{-21} \, \Big(\frac{m_{\gamma^\prime}}{10^{-20}\,{\rm eV}}\Big) \, \nonumber\\
&& \times \Big(\frac{1.26\,{\rm keV}/{\rm cm}^3}{\rho_{\rm DM}^0}\Big)^{1/2} \Big(\frac{3400}{z_{\rm eq}}\Big)^{1/2} \Big(\frac{3.94}{g_{*s}(T_{\rm eq})}\Big)^{1/2} \, . ~~~
\label{eq:BoundCosmicNu}
\eea

The medium effect at the ultrarelativistic limit can alter significantly neutrino oscillations.
Neutrino flavor oscillations hinge on the differences between the neutrino masses-squared, not the mass scale.
The consequent neutrino mass-squared differences are given by an order of $\mathcal{O}(10^{-3})$ and $\mathcal{O}(10^{-5})\,{\rm eV}^2$~\cite{deSalas:2017kay,Capozzi:2018ubv,Esteban:2018azc,Esteban:2020cvm}, whereas the mass scale involves much milder bounds that $m_\nu < \mathcal{O}(1)\,{\rm eV}$ from laboratory probes (e.g., the kinematic search in tritium decay from KATRIN~\cite{KATRIN:2019yun,KATRIN:2021uub})  and $m_\nu < \mathcal{O}(0.1)\,{\rm eV}$~\cite{Planck:2018vyg} from cosmological implications associated with the CMB spectrum and the matter power spectrum. Thus, one can put an bound of $\delta m_\nu < m_\nu$ whose precision depends on the measurement.
More importantly, if the DM contribution to the neutrino dispersion relations is lepton flavor-dependent, a tiny shift in the neutrino energy can contribute significantly to the neutrino oscillation. For instance, the solar neutrino transition is governed by the  Mikheev-Smirnov-Wolfenstein (MSW) effect~\cite{Wolfenstein:1977ue,Mikheyev:1985zog} through the effective potential induced by the solar medium. Thus the same kind of contribution from the DM, $\delta E \sim  \left| \Delta g_\nu \right|\, \delta m_\nu$ with $\Delta g_\nu = -\cos^2\theta_{13}\left(g_{\nu_e}-g_{\nu_\mu}\right)/2+\left(\sin^2\theta_{23}-\sin^2\theta_{13}\cos^2\theta_{23}\right)\left(g_{\nu_\tau}-g_{\nu_\mu}\right)/2$~\cite{Coloma:2020gfv} and $\theta_{ij}$ the mixing angles in the neutrino sector, has a significant impact on the observed (electron-)neutrino flux, and provides the stringent constraint as follows:
\bea
g_\nu  & < & \, {\rm Max}\Bigg[1.6\times 10^{-29} \, \Delta g_\nu^{-1} \Big(\frac{m_{\gamma^\prime}}{10^{-20}\,{\rm eV}}\Big)\Big(\frac{V_{\rm MSW}^\odot}{10^{-12}\,{\rm eV}}\Big) \, ,  \nonumber \\
&& \quad \,\,\,\,  6.6\times 10^{-28} \Big(\frac{m_{\gamma^\prime}}{10^{-20}\,{\rm eV}}\Big)^{3/2}\Big(\frac{m_\nu}{1\,{\rm eV}}\Big)^{1/2} \Bigg] \nonumber  \\
&&\times  \Bigg(\frac{0.3\,{\rm GeV}/{\rm cm}^{3}}{\rho^\oplus_{\rm DM}}\Bigg)^{1/2} \, ,
\label{eq:BoundNuOsc}
\eea
where $m_\nu$ denotes the neutrino mass scale in the vacuum, and ${\rm Max}[x,y]$ is the function to find the maximum value between the arguments of $x$ and $y$.
The first condition in the $\rm Max$ function is derived from the comparison of the effective neutrino potentials from the DM ($ =\sqrt{{ 1}/3}\delta m_\psi$) and from the matter  ($=\sqrt{2}G_F n_e^\odot \sim 10^{-12}\,{\rm eV}$~\cite{Bahcall:2000nu,Serenelli:2011py}  with $G_F$ the Fermi constant and $n_e^\odot$ the electron number density inside the Sun).
At the same time, the condition $\delta m_\psi^2 > m_\psi m_{\gamma^\prime}$ has to be satisfied, and thus the second constraint is imposed with $m_\nu = 1\,{\rm eV}$ as a fiducial value.

\begin{figure}
\centering
\includegraphics[width=0.475\textwidth]{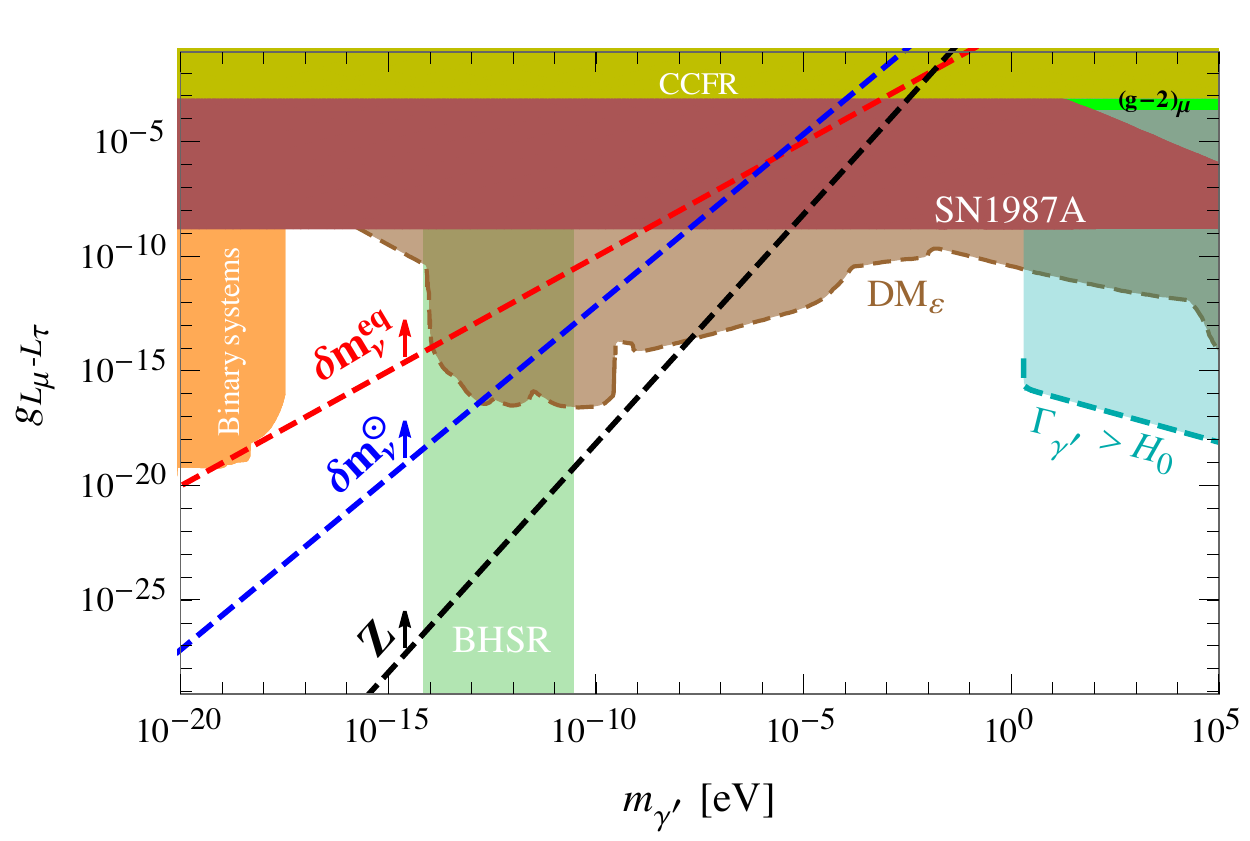}
\caption{\em The constraints plot for the gauged $L_\mu - L_\tau$ model.  The dashed red ($\delta m_\nu^{\rm eq}$), blue ($\delta m_\nu^\odot$), and black ($Z$)  lines indicate the upper bounds from Eqs.~\eqref{eq:BoundCosmicNu}, \eqref{eq:BoundNuOsc}, and \eqref{eq:Zbound}, respectively. In the cyan region ($\Gamma_{\gamma^\prime} > H_0$), dark matters are unstable. The brown region (DM$_\varepsilon$) is excluded in the presence of the natural amount of the kinetic mixing induced by the muon and tau loops~\cite{Escudero:2019gzq}. The current best bounds come from cooling of supernovae (SN1987A)~\cite{Croon:2020lrf}, from orbital period decay of compact binary systems (Binary systems), from neutrino tridents (CCFR)~\cite{CCFR:1991lpl,Altmannshofer:2014pba}, and from blackhole superradiance (BHSR)~\cite{Cardoso:2018tly,Stott:2020gjj,Ghosh:2021zuf}. The green region denoted by $(g-2)_\mu$ can address the deviation in the anomalous muon magnetic moment~\cite{Czarnecki:2001pv,Baek:2001kca,Heeck:2022znj}.}
\label{fig:LmuLtauBounds}
\end{figure}


As can be seen from Eqs.~\eqref{VZ2}-\eqref{AZ1}, the wave-function normalization factor for the fermions coupling to VDM is $1/2$ $(\delta m_\psi^2 > m_\psi m_{\gamma^\prime})$ or $3 m_{\gamma^\prime}^2/\delta m_\psi^2$ $(\delta m_\psi^2 < m_\psi m_{\gamma^\prime})$ in the relativistic limit unless $\delta m^2_\psi \ll m_{\gamma'}^2$.
This implies that lepton flavor universality can be broken badly in the gauged $L_e-L_\mu$ model, etc. On the other hand, the lepton universality, e.g., in the $Z$ boson decay is maintained at the precision of  $\sim 0.3\, \%$~\cite{Zyla:2020zbs}.
In case of $B-L$, only SM fermions interact with the VDM and thus their couplings to the SM gauge bosons  $V$-$\psi$-$\psi'$ are reduced by the normalization factor while the triboson couplings  $V$-$V$-$V$ are unaltered. Thus any deviation of these gauge couplings from the SM prediction is also tightly constrained. The triboson couplings are measured at the level of $\sim 1$ \% \cite{ALEPH:2013dgf,CMS:2019ppl}.
Finally, the dark photon with kinetic mixing has a highly suppressed coupling to neutrinos compared with the accompanying charged leptons, which spoils the $SU(2)$ doublet structure of SM. That is, the SM prediction of the ratio $\Gamma(Z\to \nu\nu)/\Gamma(Z\to ll)$ for instance could be modified significantly. Note that this ratio measures the number of neutrino species which is well determined at the precision below $1$ \%~\cite{ALEPH:2005ab}.
As the deviation of the normalization factor from unity is measure by $1-Z \approx  \delta m_\psi^2/3m_{\gamma^\prime}^2$ for $m_{\gamma^\prime}^2 \gg \delta m_\psi^2$,
one can put a quite generic bound of $\delta m_\psi < 0.1 m_{\gamma'}$ in the VDM models, which reads as follows:
\be
g_\psi < 6.6 \times 10^{-39} \Bigg(\frac{m_{\gamma^\prime}}{10^{-20}\,{\rm eV}}\Bigg)^{2}\Bigg(\frac{0.3\,{\rm GeV}/{\rm cm}^{3}}{\rho^\oplus_{\rm DM}}\Bigg)^{1/2} \, . 
\label{eq:Zbound}
\ee
This rules out a huge region of the VDM parameter space.

\begin{figure}
\centering
\includegraphics[width=0.475\textwidth]{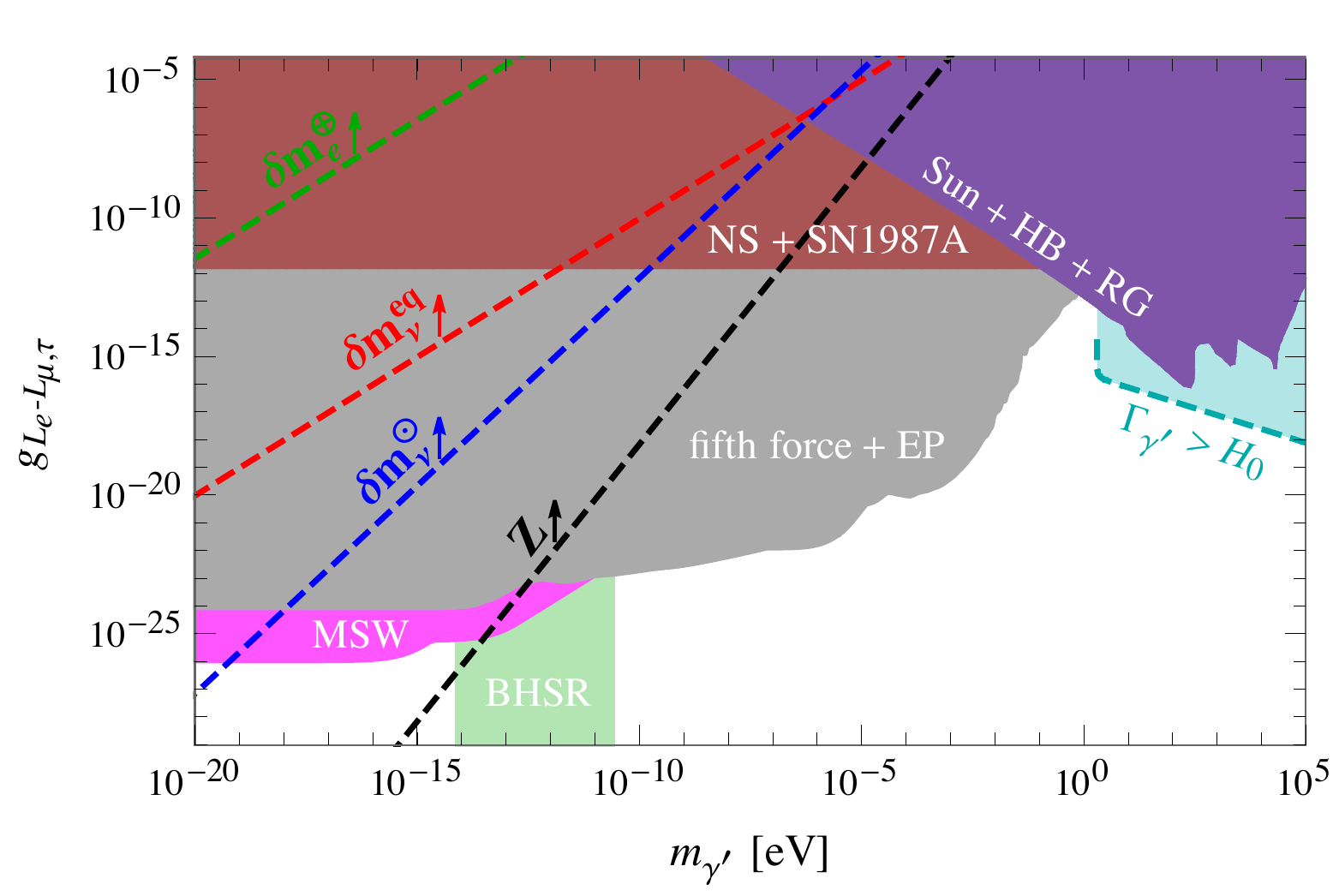}
\caption{\em The constraints plot for the gauged $L_e - L_\tau$ or $L_e - L_\tau$ model.  The dashed red ($\delta m_\nu^{\rm eq}$), green ($\delta m_e^\oplus$), blue ($\delta m_\nu^\odot$), and black ($Z$) lines indicate the upper bounds from Eqs.~\eqref{eq:BoundCosmicNu}, \eqref{eq:BoundEmass}, \eqref{eq:BoundNuOsc}, and \eqref{eq:Zbound}, respectively. In the cyan region ($\Gamma_{\gamma^\prime} > H_0$), dark matters are unstable. The current best bounds come from the gravity experiments (fifth force + EP)~\cite{Smith:1999cr,Fischbach:1999bc,Adelberger:2009zz,Wagner:2012ui,Murata:2014nra,Fayet:2017pdp}, from cooling of young neutron stars (NS+SN1987A)~\cite{Shin:2021bvz}, from cooling of the Sun, horizontal branch stars, and red giants (Sun+HB+RG)~\cite{An:2013yfc,Redondo:2013lna}, from matter effects on neutrino oscillations (MSW)~\cite{Coloma:2020gfv}, and from blackhole superradiance (BHSR)~\cite{Cardoso:2018tly,Stott:2020gjj,Ghosh:2021zuf}.}
\label{fig:LeLmutauBounds}
\end{figure}

\medskip

Figure~\ref{fig:LmuLtauBounds} shows the excluded parameter region of the gauged $L_\mu - L_\tau$ model.
No constraint comes from the experiments to determine the electron mass, but  we have the upper bounds associated with the neutrino dispersion and lepton universality; the dashed red ($\delta m_\nu^{\rm eq}$), blue ($\delta m_\nu^\odot$), and black ($Z$) lines from Eq.~\eqref{eq:BoundCosmicNu}, Eq.~\eqref{eq:BoundNuOsc}, and Eq.~\eqref{eq:Zbound}, respectively.
Furthermore, due to decay into neutrinos, the DM stability condition excludes the cyan region ($\Gamma_{\gamma^\prime} > H_0$).
The kinetic mixing of the $L_\mu-L_\tau$ gauge boson with the SM photon can be radiatively induced by the muon and tau loops, which lead to $\varepsilon = e g_{L_\mu-L_\tau} \log [m_\tau^2/m_\mu^2]/12\pi^2$~\cite{Escudero:2019gzq}.
The brown region (${\rm DM}_\varepsilon$) accounts for the additional constraints from this natural amount of the kinetic mixing, which are rescaled from the dark photon model as we will discuss at the end of this section, although those bounds possess  theoretical uncertainties that an explicit $\varepsilon$ value can be modified or even cancelled out in the UV completion.
From now on, the dashed boundaries and lines indicate a model dependence of the constraint in the assumption that the DM is fully constituted by the classically oscillating vector field.
For comparison, we also report the other constraints, which are not related to the DM.
The constraints from the gravity experiments are absent since there is no additional long-range force among baryons and electrons in nuclei; even though there is a radiatively induced kinetic mixing with the photon as discussed above, all the nuclei are electrically neutral, so still no bound from the gravity experiments is expected.
The terrestrial searches for the neutrino trident production that the $L_\mu - L_\tau$ gauge boson can mediate the interaction of muon neutrinos to heavy nuclei, which leads to $\mu^- \mu^+$ pair creation, provide the constraint (CCFR)~\cite{CCFR:1991lpl,Altmannshofer:2014pba}.
The astrophysical searches also give rise to the stringent constraints.
The stellar cooling argument~\cite{Raffelt:1996wa} on the core-collapse supernova that non-negligible muon abundance~\cite{Brust:2013ova,Bollig:2017lki} emits energetic $L_\mu - L_\tau$ gauge bosons, then such an extra energy leakage can modify the neutrino observations (SN1987A)~\cite{Croon:2020lrf} .
Furthermore, such a large muon charge of neutron stars induces a dipole $L_\mu - L_\tau$ gauge boson radiation in compact binary systems, the energy loss of which contributes to the decay of orbital period (Binary systems)~\cite{KumarPoddar:2019ceq}.
There are also the blackhole superradiance bounds~\cite{Cardoso:2018tly,Stott:2020gjj,Ghosh:2021zuf} (BHSR) on a light mass of the gauge boson that would spin down stellar mass black holes.
The green bar denoted by $\left(g-2\right)_\mu$~\cite{Czarnecki:2001pv,Baek:2001kca,Heeck:2022znj}  indicates the region to address the deviation in the anomalous muon magnetic moment~\cite{Muong-2:2021ojo}, which gives a strong motivation to the gauged $L_\mu - L_\tau$ model.

\begin{figure}
\centering
\includegraphics[width=0.475\textwidth]{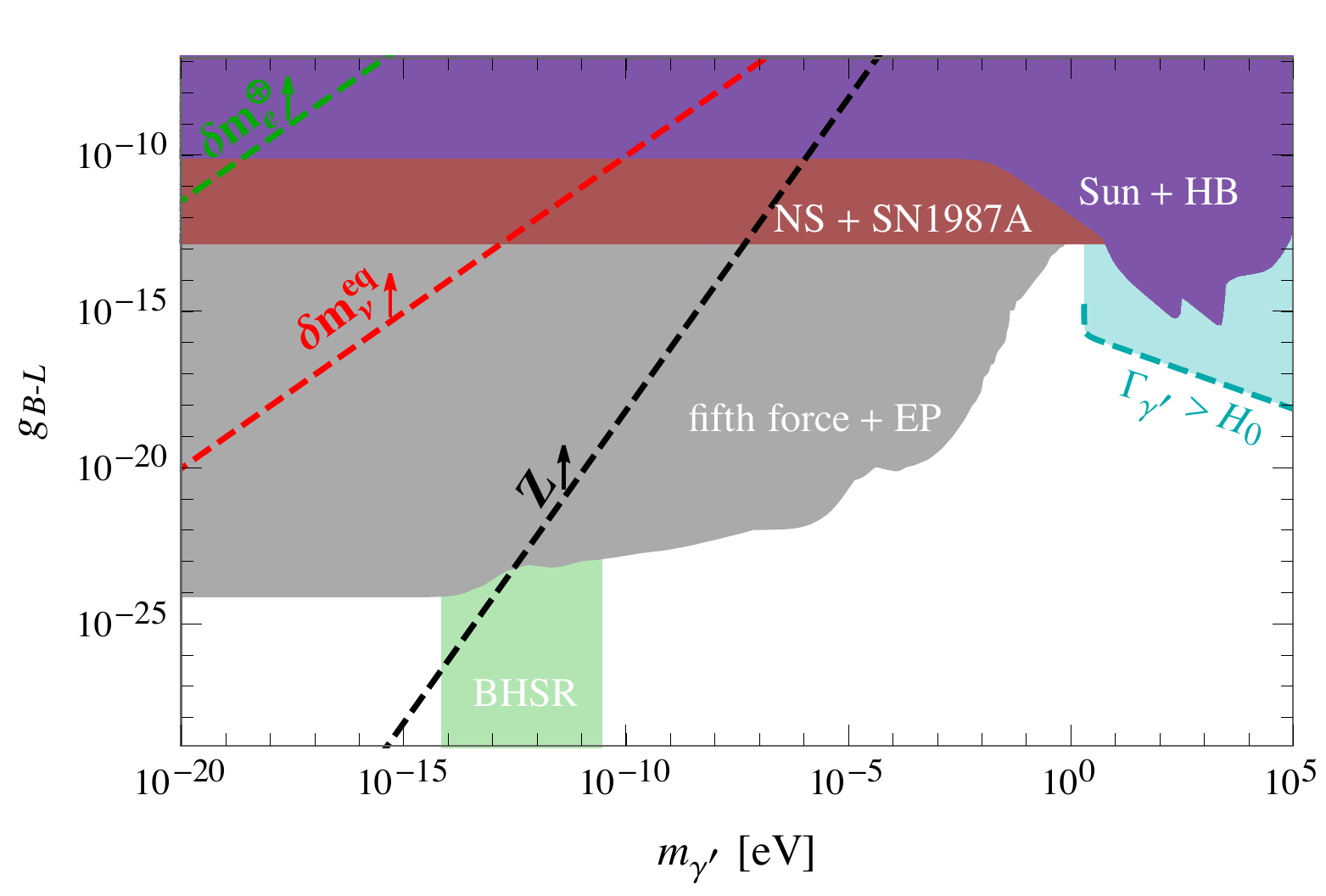}
\caption{\em The constraints plot for the gauged $B-L$ model. The dashed red ($\delta m_\nu^{\rm eq}$), green ($\delta m_e^\oplus$), and black ($Z$) lines indicate the upper bounds from Eqs.~\eqref{eq:BoundCosmicNu}, \eqref{eq:BoundEmass}, and \eqref{eq:Zbound}, respectively. In the cyan region ($\Gamma_{\gamma^\prime} > H_0$), dark matters are unstable. The current best bounds come from the gravity experiments (fifth force + EP)~\cite{Smith:1999cr,Fischbach:1999bc,Adelberger:2009zz,Wagner:2012ui,Murata:2014nra,Fayet:2017pdp}, from cooling of young neutron stars (NS+SN1987A)~\cite{Shin:2021bvz}, from cooling of the Sun and horizontal branch stars (Sun+HB)~\cite{Hardy:2016kme}, and from blackhole superradiance (BHSR)~\cite{Cardoso:2018tly,Stott:2020gjj,Ghosh:2021zuf}.}
\label{fig:BLBounds}
\end{figure}

In the gauged $L_e - L_{\mu}$ or $L_e - L_\tau$ model, we can access the constraints from the electron dispersion in the experiments as well as implications of the neutrino mass scale.
In Fig.~\ref{fig:LeLmutauBounds},  the dashed red ($\delta m_\nu^{\rm eq}$), green ($\delta m_e^\oplus$), blue ($\delta m_\nu^\odot$', and black ($Z$) lines illustrate the corresponding upper bounds of Eqs.~\eqref{eq:BoundCosmicNu}, \eqref{eq:BoundEmass}, \eqref{eq:BoundNuOsc}, and \eqref{eq:Zbound}, respectively.
The cyan region accounts for the DM stability bound ($\Gamma_{\gamma^\prime} > H_0$) as in the gauged $L_\mu - L_\tau$ model.
The gravity experiments for tests of the equivalence principle~\cite{Smith:1999cr,Wagner:2012ui,Fayet:2017pdp} and fifth forces~\cite{Fischbach:1999bc,Adelberger:2009zz,Murata:2014nra} come up with the most stringent bound for masses below an order of $\rm eV$ (fifth force + EP).
Likewise, a long-range mediator induces a nonstandard matter potential in neutrino propagation, which would modify the oscillation data (MSW)~\cite{Coloma:2020gfv}.
For the astrophysical bounds, we can exploit the known results in the other dark gauge boson models.
In less dense stars such as the Sun, horizontal branch stars, and red giants, the main emission process is the resonant conversion of longitudinal plasmons~\cite{An:2013yfc,Redondo:2013lna}, which is supported by the in-medium mixing~\cite{Hardy:2016kme,Hong:2020bxo}, thus the constraints are rescaled from the result in the dark photon model (Sun+HB+RG).
The other significant astrophysical constraints emerge from the cooling observations of neutron stars (NS+SN1987A)~\cite{Shin:2021bvz} where the nucleon bremsstrahlung is typically the dominant production channel; the leading order emissivity is controlled by the coupling difference between the nucleon fields (as the isospin breaking factor~\cite{Shin:2021bvz}), which is given equivalently in both the gauged $U(1)$ extensions of ${L_e}-L_{\mu,\tau}$ and ${B-L}$ number on account of the in-medium effect~\cite{Hardy:2016kme,Hong:2020bxo}.
We include the blackhole superradiance bounds (BHSR)~\cite{Cardoso:2018tly,Stott:2020gjj,Ghosh:2021zuf}.

As shown in Fig.~\ref{fig:BLBounds}, most of the constraints on the gauged $B-L$ model resemble  the gauged $L_e-L_{\mu , \tau}$ models, but those from neutrino oscillations (i.e. the constraints denoted by $\delta m_\nu^\odot$ and MSW in Fig.~\ref{fig:LeLmutauBounds}) are absent due to the flavor-universal aspect of the gauge charge assignment.
Furthermore, we report a difference in the astrophysical constraints from low density stellar objects (Sun  + HB) that becomes flat for masses below $10^{-2}\,{\rm eV}$~\cite{Hardy:2016kme} due to the contribution from neutrons, the coupling of which experience no in-medium effect.

\begin{figure}
\centering
\includegraphics[width=0.475\textwidth]{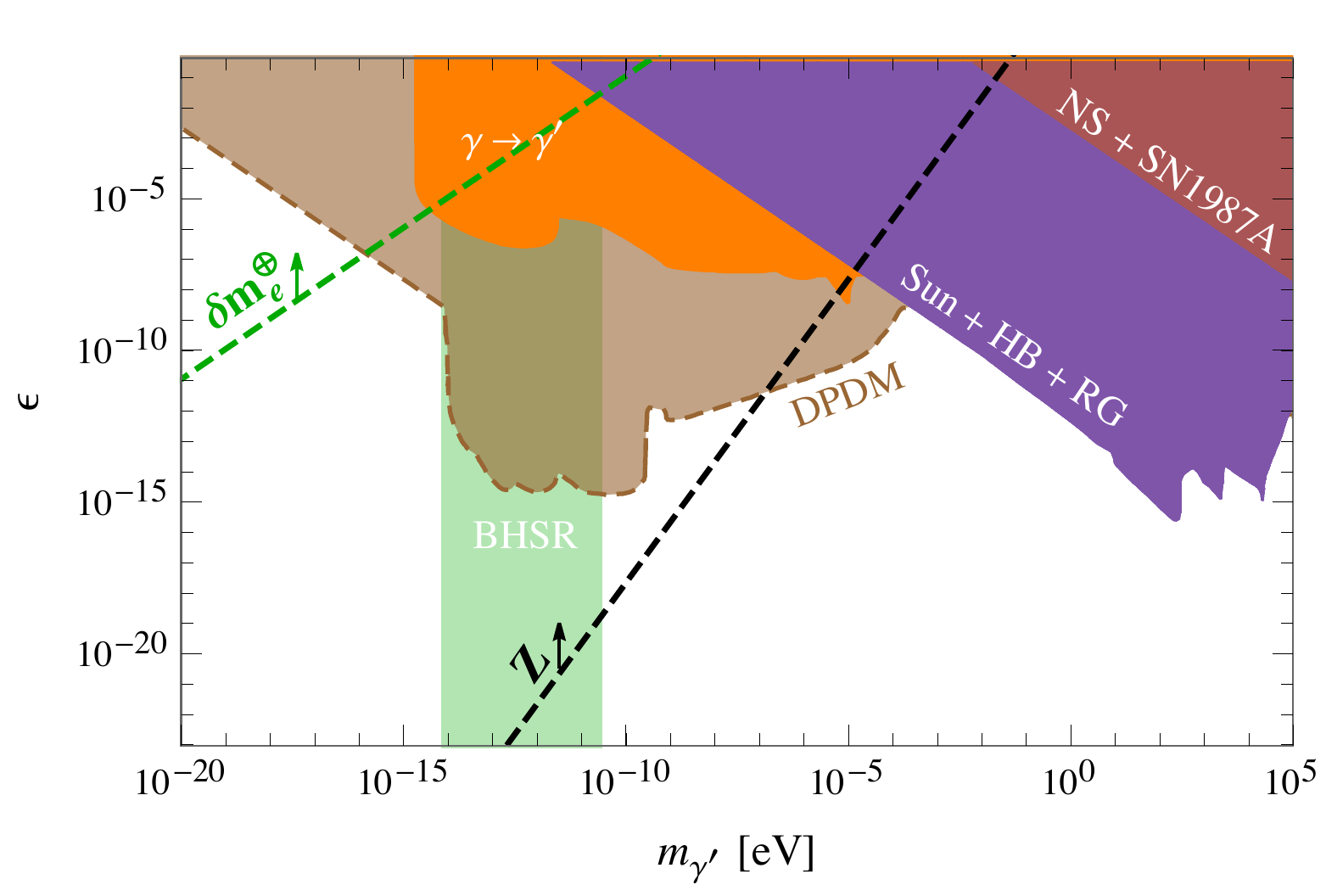}
\caption{\em The constraints plot for the dark photon model.  The dashed green ($\delta m_e^\oplus$) and black ($Z$) lines indicate the upper bounds from Eqs.~\eqref{eq:BoundEmass} and \eqref{eq:Zbound}, respectively. The brown region (DPDM) corresponds to the preexisting constraints on the dark photon DM~\cite{Arias:2012az,Dubovsky:2015cca,Wadekar:2019mpc,Bhoonah:2019eyo,McDermott:2019lch,Caputo:2020bdy,Witte:2020rvb,Caputo:2020rnx,Caputo:2021eaa}. The current best bounds come from cooling of young neutron stars (NS+SN1987A)~\cite{Chang:2016ntp,Shin:2021bvz}, from cooling of the Sun, horizontal branch stars, and red giants (Sun+HB+RG)~\cite{An:2013yfc,Redondo:2013lna}, from transitions of SM photons to dark photons ($\gamma\to\gamma^\prime$)~\cite{Povey:2010hs,Ehret:2010mh,ADMX:2010ubl,Inada:2013tx,Betz:2013dza,Parker:2013fxa,Caputo:2020bdy}, and from blackhole superradiance (BHSR)~\cite{Cardoso:2018tly,Stott:2020gjj,Ghosh:2021zuf}.}
\label{fig:DPBounds}
\end{figure}

The last model that we explore is the dark photon scenario, where the kinetic mixing with the SM photon induces the leading order couplings to the SM particles.
The constraints for the dark photon model are depicted in Fig.~\ref{fig:DPBounds}.
Since the dark photon couples only to SM particles containing a nonvanishing electric charge,  we can only demand the upper bound of Eqs.~\eqref{eq:BoundEmass} and \eqref{eq:Zbound} from the electron mass measurements and the electroweak precision tests, which correspond to the dashed green ($\delta m_e^\oplus$) and black ($Z$) lines in Fig.~\ref{fig:DPBounds}, respectively.
The kinetic mixing transfers a fractional energy density of the VDM into cosmological and astrophysical mediums.
Such an energy deposition could alter the observations as a heating source, then we can obtain various constraints on the dark photon DM (see Ref.~\cite{Caputo:2021eaa} and references therein~\cite{Arias:2012az,Dubovsky:2015cca,Wadekar:2019mpc,Bhoonah:2019eyo,McDermott:2019lch,Caputo:2020bdy,Witte:2020rvb,Caputo:2020rnx} for details).
Taking the most stringent constraint for each dark photon mass, the brown region with the dashed boundary (DPDM) is excluded when the coherent dark photon field constitutes the whole DM abundance; the rescaled bounds are used in the gauged $L_\mu - L_\tau$ model as shown in Fig.~\ref{fig:LmuLtauBounds}.
Apart from the constraints above in terms of the DM that are based on the conversion of dark photons into SM photons, there are also the constraints that rely on transitions in the opposite way ($\gamma\to \gamma^\prime$) and are less model-dependent; as an example, light-shining-through-walls experiments~\cite{Povey:2010hs,Ehret:2010mh,ADMX:2010ubl,Inada:2013tx,Betz:2013dza,Parker:2013fxa} and CMB spectral distortions~\cite{Caputo:2020bdy}.
At masses above $10^{-3}\,{\rm eV}$, the stellar cooling argument on the Sun, horizontal branch stars, red giants (Sun+HB+RG)~\cite{An:2013yfc,Redondo:2013lna}, neutron stars, and supernovae (NS+SN1987A)~\cite{Chang:2016ntp,Shin:2021bvz} gives the best current bounds.
We also include the model-independent constraints from blackhole superradiance (BHSR)~\cite{Cardoso:2018tly,Stott:2020gjj,Ghosh:2021zuf}.

\section{Conclusions and discussion}
\label{sec:conclusions}

We discussed the medium effect on the particles living in the classical vector field background which is considered as a light DM candidate.
Even though the DM polarization within at least a local domain can be aligned with a single direction or randomly distributed, its impacts on the particle dispersion would be washed out by integrating out uncertainties associated with a particle momentum in the DM frame.
We computed a degree of the modification of the particle dispersion in each limited circumstance, which is quantified by $\Delta$ in Eq.~\eqref{eq:Delta}.
In the case of a vector gauge charge assignment, as long as the gauge coupling is large to achieve the condition of $\delta m_\psi^2 \gg m_\psi m_{\gamma^\prime}$, such a modification becomes significant due to the enhancement from the intermediate particle mediator.
Based on these findings, we  evaluated the constraints on the gauged $U(1)$ models of interest and the results are depicted in Figs.~\ref{fig:LmuLtauBounds}, \ref{fig:LeLmutauBounds}, \ref{fig:BLBounds}, and \ref{fig:DPBounds} for the gauged $L_\mu - L_\tau$, $L_e-L_{\mu,\tau}$, $B-L$, and dark photon models, respectively.
Interestingly, the DM implications of refractive phenomena give the stringent constraints on the gauged $L_\mu-L_\tau$ and dark photon models, where there is no bound from the gravity experiments due to no effective baryonic coupling to nuclei.

Furthermore, we pointed out that the effective coupling strengths in the relativistic regime could deviate from unity, which leads to a conflict with the SM prediction.
The electroweak precision tests indeed impose a severe constraint on the VDM with ultralight masses of $m_{\gamma^\prime}< 10^{-15}\,{\rm eV}$.

Besides the discussion of cosmological neutrino refraction and its implications, one may wonder the constraints from the particle dispersion of electrons (as lightest charged particles) in cosmology.
A large correction to the electron mass in the early universe would be problematic for the cosmological history.
For example, the change of the recombination epoch might be expected due to the modified binding energy of hydrogen as a naive inference.
In order to account for the DM effect on the dispersion relation of electrons, we need to deal with the plasma screening effect in the ionized medium, then the effective DM coupling to electrons can be written by (for details, see Refs.~\cite{Hardy:2016kme,Hong:2020bxo})
\bea
g_e^{\rm eff} = \left( \varepsilon e - g_e \right)\frac{m_{\gamma^\prime}^2}{m_{\gamma}^2 - m_{\gamma^\prime}^2} \, .
\label{eq:eEffCoup}
\eea
Here $m_{\gamma}^2$ quantifies an effective photon mass with the imaginary part, which accounts for photon absorption (for details, see Ref.~\cite{Arias:2012az} and references therein~\cite{Ahonen:1995ky,Mirizzi:2009iz}).
Following Ref.~\cite{Arias:2012az}, we notice that the constraint from the electron refraction at the recombination  is much weaker than the results above in spite of the resonance in Eq.~\eqref{eq:eEffCoup} for $m_{\gamma^\prime}\simeq 10^{-9}\,{\rm eV}$  at the recombination.

We ignore any bounds from the big bang nucleosynthesis (BBN) era, which are rather large model-dependent.
When the Hubble friction dominates over the DM mass rolling in the potential, the DM field is stuck at the initial value and its density scales as constant, not $\rho \propto a^{-3}$ of the oscillating field.
Therefore, if $m_{\gamma^\prime} < H_{\rm BBN} \sim \mathcal{O}(10^{-16})\,{\rm eV}$, the modification of the particle dispersion against the vacuum at the BBN depends on the initial misalignment, and the derivation discussed in Sec.~\ref{sec:Dispersion} is not valid any longer.
Furthermore, we can imagine that the DM is generated after the BBN, which is not in conflict with the standard cosmological evolution.

As a final remark, it would be interesting to explore any phenomenological aspects of the DM polarization.
Indeed, the DM polarization is responsible for distinct features in direct detection searches compared to axions, and a careful analysis to point out the validity of the historical reinterpretation of axion bounds is carried out in the literature~\cite{Arias:2012az,Caputo:2021eaa}.
As discussed in Sec.~\ref{sec:unpol} and Appendix, particle dispersions can also depend on the polarization, thus it might leave an imprint on observations in terrestrial or cosmological experiments.

\begin{acknowledgments}
Authors acknowledge S. Bhattacharya, F. D'Eramo, C. S. Shin, and L. X. Xu for useful discussions. This work is supported by the research grants: ``The Dark Universe: A Synergic Multi-messenger Approach'' number 2017X7X85K under the program PRIN 2017 funded by the Ministero dell'Istruzione, Universit\`a e della Ricerca (MIUR); ``New Theoretical Tools for Axion Cosmology'' under the Supporting TAlent in ReSearch@University of Padova (STARS@UNIPD). S.Y. also supported by Istituto Nazionale di Fisica Nucleare (INFN) through the Theoretical Astroparticle Physics (TAsP) project.
\end{acknowledgments}

\appendix

\section{$\Delta$ in the polarized background}

The equation of motion in Eq.~\eqref{eq:AXEoM} leads to the quintic equation of $\Delta$ as follows
\bea
\sum_{i=0}^5 \tilde{\Upsilon}_i \Delta^i = 0 \,
\label{eq:DeltaEqUnPol}
\eea
with
\bea
\tilde{\Upsilon}_0 & = &m_{\gamma^\prime}^4 \delta m_\psi^2 \Bigg[ \left(1-\alpha_m^2\right)  m_\psi^2 \delta m_\psi^2  +\Big(\frac{m_{\gamma^\prime}^2}{4}- m_\psi^2 - |\vec{p}|^2\Big) \nonumber \\
&& \times 16 \Big(\sin^2\theta |\vec{p}|^2 + \frac{1+ \alpha_m}{2}m_\psi^2\Big) \Bigg] \, , \\
\tilde{\Upsilon}_1 &=& m_{\gamma^\prime}^2 \Bigg[-2\delta m_\psi^2 \Big(\delta m_\psi^2 \left(2 |\vec{p}|^2+\left(1+\alpha_m^2\right) m_\psi^2 \right) \nonumber \\
&& -4\left(m_\psi^2+|\vec{p}|^2\right)\left(2\cos^2\theta|\vec{p}|^2+\left(1-\alpha_m\right) m_\psi^2\right)\Big) \nonumber   \\
&& +m_{\gamma^\prime}^2\Bigg( \delta m_\psi^4 + 16\left(m_\psi^2+|\vec{p}|^2\right)^2 \nonumber \\
&&-4\delta m_\psi^2 \left(\left(4-\cos^2\theta\right) |\vec{p}|^2 + \frac{7+\alpha_m}{2}m_\psi^2\right) \Bigg) \nonumber \\
&&  + 2 m_{\gamma^\prime}^4 \Big(\delta m_\psi^2 -4 m_\psi^2 -4 |\vec{p}|^2\Big)+m_{\gamma^\prime}^6\Bigg] \, , \\
\tilde{\Upsilon}_2 &=&  \delta m_\psi^4 m_\psi^2  \left(1-\alpha_m^2\right) + 4m_{\gamma^\prime}^4 \Big(4 \left(m_\psi^2+|\vec{p}|^2\right)-\frac{3}{2} \delta m_\psi^2\Big) \nonumber\\
&&  - 2\delta m_\psi^2 m_{\gamma^\prime}^2 \Big(\delta m_\psi^2-\left(7-\alpha_m\right) m_\psi^2 \nonumber\\
&&- 2\left(3+\cos^2\theta\right) |\vec{p}|^2\Big)   -4 m_{\gamma^\prime}^6 \, ,\\
\tilde{\Upsilon}_3 &= &  \delta m_\psi^4 - \delta m_\psi^2\Big( 4\cos^2\theta |\vec{p}|^2 + 2\left(1- \alpha_m\right)  m_\psi^2\Big) \nonumber \\
&& +m_{\gamma^\prime}^2 \Big(6\delta m_\psi^2-8\left(m_\psi^2 +  |\vec{p}|^2\right)\Big) +6 m_{\gamma^\prime}^4   \, , \\
\tilde{\Upsilon}_4 &=& -2 \delta m_\psi^2 -4 m_{\gamma^\prime}^2  \, , \qquad \tilde{\Upsilon}_5 = 1 \,
\eea
with the definition of $\cos\theta = \hat{z}\cdot \vec{p}/|\vec{p}|$.

 Let us derive the solution of $\Delta$ in a few limited conditions that cover most of our interest.
The proper $\Delta$ values in the polarized background have dependence of $\cos\theta$.
We take the assumption of $m_{\gamma^\prime}^2 \ll m_\psi^2\, , \delta m_\psi^2$ for convenience.
\smallskip

\underline{\it Vector coupling}:

The comparison between $\delta m_\psi^2 \cos^2\theta$ and $m_{\gamma^\prime}$ as well as between $\delta m_\psi^2$ and $m_\psi m_{\gamma^\prime}$ characterize the behavior of the solutions.
In the typical range of $\cos^2\theta$ for $\delta m_\psi^2 \cos^2\theta \gg m_{\gamma^\prime}^2$, we obtain the results for the small coupling regime of $\delta m_\psi^2 \ll m_\psi m_{\gamma^\prime}$
\begin{equation}
\Delta \approx \left\{
\begin{split}
& \frac{\delta m_\psi^2 m_{\gamma^\prime}^2 \sin^2\theta}{\delta m_\psi^2 \cos^2\theta +  m_{\gamma^\prime}^2} & \mbox{for}~~ \sqrt{\sin^2\theta}|\vec{p}|\gg  m_\psi \\
& \delta m_\psi^2 & \mbox{for}~~ \sqrt{\sin^2\theta}|\vec{p}|\ll  m_\psi
\end{split}
\right. \, ,
\label{eq:UnPolVec1}
\end{equation}
and for the opposite regime of $\delta m_\psi^2 \gg m_\psi m_{\gamma^\prime}$
\begin{equation}
\Delta \approx \left\{
\begin{split}
& 2 \sqrt{\cos^2\theta } |\vec{p}| \delta m_\psi & \mbox{for}~~  \sqrt{\cos^2\theta} |\vec{p}|\gg  \delta m_\psi \\
& \delta m_\psi^2 & \mbox{for}~~ \sqrt{\cos^2\theta} |\vec{p}|\ll  \delta m_\psi
\end{split}
\right. \, .
\label{eq:UnPolVec2}
\end{equation}
In the limit of $\delta m_\psi^2 \cos^2\theta \ll m_{\gamma^\prime}^2$, the $\Delta$ solution for $\delta m_\psi^2 \ll m_\psi m_{\gamma^\prime}$ is equivalent to Eq.~\eqref{eq:UnPolVec1}, whereas the result for $\delta m_\psi^2 \gg m_\psi m_{\gamma^\prime}$ is still similar to Eq.~\eqref{eq:UnPolVec1} but with the difference kinematic condition as following
\begin{equation}
\Delta \approx \left\{
\begin{split}
&\frac{\delta m_\psi^2 m_{\gamma^\prime}^2 \sin^2\theta}{\delta m_\psi^2 \cos^2\theta +  m_{\gamma^\prime}^2} & \mbox{for}~~ |\vec{p}|\gg  \delta m_\psi^2/m_{\gamma^\prime} \\
& \delta m_\psi^2 & \mbox{for}~~ |\vec{p}|\ll  \delta m_\psi^2/m_{\gamma^\prime}
\end{split}
\right. \, .
\label{eq:UnPolVec3}
\end{equation}
\smallskip

\underline{\it Axial-vector coupling}:

The common solutions for $\delta m_\psi^2 \ll m_\psi^2$ are given by
\begin{equation}
\Delta \approx \left\{
\begin{split}
& \frac{\delta m_\psi^2 m_{\gamma^\prime}^2 \sin^2\theta}{\delta m_\psi^2 \cos^2\theta +  m_{\gamma^\prime}^2} & \mbox{for}~~ |\vec{p}|\gg  \mathcal{F} m_\psi \\
& \sin^2\theta\frac{m_{\gamma^\prime}^2\delta m_\psi^2}{m_{\gamma^\prime}^2+\delta m_\psi^2}\frac{|\vec{p}|^2}{m_\psi^2} & \mbox{for}~~ |\vec{p}|\ll  \mathcal{F} m_\psi
\end{split}
\right. \,
\label{eq:UnPolAX1}
\end{equation}
with $\mathcal{F} = \left(\sqrt{\cos^{2}\theta}\right)^{-1}$ and $\sqrt{1+\delta m_\psi^2/m_{\gamma^\prime}^2}$ in the regime of $\delta m_\psi^2 \cos^2\theta \gg m_{\gamma^\prime}^2$ and $\delta m_\psi^2 \cos^2\theta \ll m_{\gamma^\prime}^2$, respectively.
As in Eq.~\eqref{Asol2}, we find the solutions, which are relevant for the $\delta m_\psi^2 \gg m_{\gamma^\prime}^2$ case and also cover the $\delta m_\psi^2 \gg m_\psi^2$ limit, as follows
\begin{equation}
\Delta \approx \left\{
\begin{split}
& 2 \sqrt{\cos^2\theta}  |\vec{p}| \delta m & \mbox{for}~~  |\vec{p}|\gg \frac{ \sqrt{m_\psi^2+\delta m_\psi^2} }{\sqrt{\cos^2\theta}} \\
& 2\delta m_\psi m_\psi + \delta m_\psi^2 & \mbox{for}~~  |\vec{p}|\ll \frac{ \sqrt{m_\psi^2+\delta m_\psi^2} }{\sqrt{\cos^2\theta}}
\end{split}
\right. \,
\label{eq:UnPolAX2}
\end{equation}
in the regime of $\delta m_\psi^2 \cos^2\theta \gg m_{\gamma^\prime}^2$ and the same result in Eq.~\eqref{eq:UnPolAX1} in the opposite regime of $\delta m_\psi^2 \cos^2\theta \ll m_{\gamma^\prime}^2$.

\bibliographystyle{apsrev4-2}
\bibliography{DispersionVDM}

\begin{thebibliography}{116}%
\makeatletter
\providecommand \@ifxundefined [1]{%
 \@ifx{#1\undefined}
}%
\providecommand \@ifnum [1]{%
 \ifnum #1\expandafter \@firstoftwo
 \else \expandafter \@secondoftwo
 \fi
}%
\providecommand \@ifx [1]{%
 \ifx #1\expandafter \@firstoftwo
 \else \expandafter \@secondoftwo
 \fi
}%
\providecommand \natexlab [1]{#1}%
\providecommand \enquote  [1]{``#1''}%
\providecommand \bibnamefont  [1]{#1}%
\providecommand \bibfnamefont [1]{#1}%
\providecommand \citenamefont [1]{#1}%
\providecommand \href@noop [0]{\@secondoftwo}%
\providecommand \href [0]{\begingroup \@sanitize@url \@href}%
\providecommand \@href[1]{\@@startlink{#1}\@@href}%
\providecommand \@@href[1]{\endgroup#1\@@endlink}%
\providecommand \@sanitize@url [0]{\catcode `\\12\catcode `\$12\catcode
  `\&12\catcode `\#12\catcode `\^12\catcode `\_12\catcode `\%12\relax}%
\providecommand \@@startlink[1]{}%
\providecommand \@@endlink[0]{}%
\providecommand \url  [0]{\begingroup\@sanitize@url \@url }%
\providecommand \@url [1]{\endgroup\@href {#1}{\urlprefix }}%
\providecommand \urlprefix  [0]{URL }%
\providecommand \Eprint [0]{\href }%
\providecommand \doibase [0]{https://doi.org/}%
\providecommand \selectlanguage [0]{\@gobble}%
\providecommand \bibinfo  [0]{\@secondoftwo}%
\providecommand \bibfield  [0]{\@secondoftwo}%
\providecommand \translation [1]{[#1]}%
\providecommand \BibitemOpen [0]{}%
\providecommand \bibitemStop [0]{}%
\providecommand \bibitemNoStop [0]{.\EOS\space}%
\providecommand \EOS [0]{\spacefactor3000\relax}%
\providecommand \BibitemShut  [1]{\csname bibitem#1\endcsname}%
\let\auto@bib@innerbib\@empty
\bibitem [{\citenamefont {Fayet}(1980)}]{Fayet:1980ad}%
  \BibitemOpen
  \bibfield  {author} {\bibinfo {author} {\bibfnamefont {P.}~\bibnamefont
  {Fayet}},\ }\href {https://doi.org/10.1016/0370-2693(80)90488-8} {\bibfield
  {journal} {\bibinfo  {journal} {Phys. Lett. B}\ }\textbf {\bibinfo {volume}
  {95}},\ \bibinfo {pages} {285} (\bibinfo {year} {1980})}\BibitemShut
  {NoStop}%
\bibitem [{\citenamefont {Fayet}(1981)}]{Fayet:1980rr}%
  \BibitemOpen
  \bibfield  {author} {\bibinfo {author} {\bibfnamefont {P.}~\bibnamefont
  {Fayet}},\ }\href {https://doi.org/10.1016/0550-3213(81)90122-X} {\bibfield
  {journal} {\bibinfo  {journal} {Nucl. Phys. B}\ }\textbf {\bibinfo {volume}
  {187}},\ \bibinfo {pages} {184} (\bibinfo {year} {1981})}\BibitemShut
  {NoStop}%
\bibitem [{\citenamefont {Holdom}(1986)}]{Holdom:1985ag}%
  \BibitemOpen
  \bibfield  {author} {\bibinfo {author} {\bibfnamefont {B.}~\bibnamefont
  {Holdom}},\ }\href {https://doi.org/10.1016/0370-2693(86)91377-8} {\bibfield
  {journal} {\bibinfo  {journal} {Phys. Lett. B}\ }\textbf {\bibinfo {volume}
  {166}},\ \bibinfo {pages} {196} (\bibinfo {year} {1986})}\BibitemShut
  {NoStop}%
\bibitem [{\citenamefont {Fayet}(1990)}]{Fayet:1990wx}%
  \BibitemOpen
  \bibfield  {author} {\bibinfo {author} {\bibfnamefont {P.}~\bibnamefont
  {Fayet}},\ }\href {https://doi.org/10.1016/0550-3213(90)90381-M} {\bibfield
  {journal} {\bibinfo  {journal} {Nucl. Phys. B}\ }\textbf {\bibinfo {volume}
  {347}},\ \bibinfo {pages} {743} (\bibinfo {year} {1990})}\BibitemShut
  {NoStop}%
\bibitem [{Pro(2012)}]{Proceedings:2012ulb}%
  \BibitemOpen
  \href {https://doi.org/10.2172/1042577} {\emph {\bibinfo {title}
  {{Fundamental Physics at the Intensity Frontier}}}}\ (\bibinfo {year}
  {2012})\ \Eprint {https://arxiv.org/abs/1205.2671} {arXiv:1205.2671 [hep-ex]}
  \BibitemShut {NoStop}%
\bibitem [{\citenamefont {Essig}\ \emph {et~al.}(2013)\citenamefont {Essig}
  \emph {et~al.}}]{Essig:2013lka}%
  \BibitemOpen
  \bibfield  {author} {\bibinfo {author} {\bibfnamefont {R.}~\bibnamefont
  {Essig}} \emph {et~al.},\ }in\ \href@noop {} {\emph {\bibinfo {booktitle}
  {{Community Summer Study 2013}: {Snowmass on the Mississippi}}}}\ (\bibinfo
  {year} {2013})\ \Eprint {https://arxiv.org/abs/1311.0029} {arXiv:1311.0029
  [hep-ph]} \BibitemShut {NoStop}%
\bibitem [{\citenamefont {Raggi}\ and\ \citenamefont
  {Kozhuharov}(2015)}]{Raggi:2015yfk}%
  \BibitemOpen
  \bibfield  {author} {\bibinfo {author} {\bibfnamefont {M.}~\bibnamefont
  {Raggi}}\ and\ \bibinfo {author} {\bibfnamefont {V.}~\bibnamefont
  {Kozhuharov}},\ }\href {https://doi.org/10.1393/ncr/i2015-10117-9} {\bibfield
   {journal} {\bibinfo  {journal} {Riv. Nuovo Cim.}\ }\textbf {\bibinfo
  {volume} {38}},\ \bibinfo {pages} {449} (\bibinfo {year} {2015})}\BibitemShut
  {NoStop}%
\bibitem [{\citenamefont {Deliyergiyev}(2016)}]{Deliyergiyev:2015oxa}%
  \BibitemOpen
  \bibfield  {author} {\bibinfo {author} {\bibfnamefont {M.~A.}\ \bibnamefont
  {Deliyergiyev}},\ }\href {https://doi.org/10.1515/phys-2016-0034} {\bibfield
  {journal} {\bibinfo  {journal} {Open Phys.}\ }\textbf {\bibinfo {volume}
  {14}},\ \bibinfo {pages} {281} (\bibinfo {year} {2016})},\ \Eprint
  {https://arxiv.org/abs/1510.06927} {arXiv:1510.06927 [hep-ph]} \BibitemShut
  {NoStop}%
\bibitem [{\citenamefont {Alekhin}\ \emph {et~al.}(2016)\citenamefont {Alekhin}
  \emph {et~al.}}]{Alekhin:2015byh}%
  \BibitemOpen
  \bibfield  {author} {\bibinfo {author} {\bibfnamefont {S.}~\bibnamefont
  {Alekhin}} \emph {et~al.},\ }\href
  {https://doi.org/10.1088/0034-4885/79/12/124201} {\bibfield  {journal}
  {\bibinfo  {journal} {Rept. Prog. Phys.}\ }\textbf {\bibinfo {volume} {79}},\
  \bibinfo {pages} {124201} (\bibinfo {year} {2016})},\ \Eprint
  {https://arxiv.org/abs/1504.04855} {arXiv:1504.04855 [hep-ph]} \BibitemShut
  {NoStop}%
\bibitem [{\citenamefont {Curciarello}(2016)}]{Curciarello:2016jbz}%
  \BibitemOpen
  \bibfield  {author} {\bibinfo {author} {\bibfnamefont {F.}~\bibnamefont
  {Curciarello}},\ }\href {https://doi.org/10.1051/epjconf/201611801008}
  {\bibfield  {journal} {\bibinfo  {journal} {EPJ Web Conf.}\ }\textbf
  {\bibinfo {volume} {118}},\ \bibinfo {pages} {01008} (\bibinfo {year}
  {2016})}\BibitemShut {NoStop}%
\bibitem [{\citenamefont {Alexander}\ \emph {et~al.}(2016)\citenamefont
  {Alexander} \emph {et~al.}}]{Alexander:2016aln}%
  \BibitemOpen
  \bibfield  {author} {\bibinfo {author} {\bibfnamefont {J.}~\bibnamefont
  {Alexander}} \emph {et~al.}\ }(\bibinfo {year} {2016})\ \Eprint
  {https://arxiv.org/abs/1608.08632} {arXiv:1608.08632 [hep-ph]} \BibitemShut
  {NoStop}%
\bibitem [{\citenamefont {Beacham}\ \emph {et~al.}(2020)\citenamefont {Beacham}
  \emph {et~al.}}]{Beacham:2019nyx}%
  \BibitemOpen
  \bibfield  {author} {\bibinfo {author} {\bibfnamefont {J.}~\bibnamefont
  {Beacham}} \emph {et~al.},\ }\href {https://doi.org/10.1088/1361-6471/ab4cd2}
  {\bibfield  {journal} {\bibinfo  {journal} {J. Phys. G}\ }\textbf {\bibinfo
  {volume} {47}},\ \bibinfo {pages} {010501} (\bibinfo {year} {2020})},\
  \Eprint {https://arxiv.org/abs/1901.09966} {arXiv:1901.09966 [hep-ex]}
  \BibitemShut {NoStop}%
\bibitem [{\citenamefont {Fabbrichesi}\ \emph {et~al.}(2020)\citenamefont
  {Fabbrichesi}, \citenamefont {Gabrielli},\ and\ \citenamefont
  {Lanfranchi}}]{Fabbrichesi:2020wbt}%
  \BibitemOpen
  \bibfield  {author} {\bibinfo {author} {\bibfnamefont {M.}~\bibnamefont
  {Fabbrichesi}}, \bibinfo {author} {\bibfnamefont {E.}~\bibnamefont
  {Gabrielli}},\ and\ \bibinfo {author} {\bibfnamefont {G.}~\bibnamefont
  {Lanfranchi}}\ }\href {https://doi.org/10.1007/978-3-030-62519-1}
  {10.1007/978-3-030-62519-1} (\bibinfo {year} {2020}),\ \Eprint
  {https://arxiv.org/abs/2005.01515} {arXiv:2005.01515 [hep-ph]} \BibitemShut
  {NoStop}%
\bibitem [{\citenamefont {Caputo}\ \emph {et~al.}(2021)\citenamefont {Caputo},
  \citenamefont {Millar}, \citenamefont {O'Hare},\ and\ \citenamefont
  {Vitagliano}}]{Caputo:2021eaa}%
  \BibitemOpen
  \bibfield  {author} {\bibinfo {author} {\bibfnamefont {A.}~\bibnamefont
  {Caputo}}, \bibinfo {author} {\bibfnamefont {A.~J.}\ \bibnamefont {Millar}},
  \bibinfo {author} {\bibfnamefont {C.~A.~J.}\ \bibnamefont {O'Hare}},\ and\
  \bibinfo {author} {\bibfnamefont {E.}~\bibnamefont {Vitagliano}},\ }\href
  {https://doi.org/10.1103/PhysRevD.104.095029} {\bibfield  {journal} {\bibinfo
   {journal} {Phys. Rev. D}\ }\textbf {\bibinfo {volume} {104}},\ \bibinfo
  {pages} {095029} (\bibinfo {year} {2021})},\ \Eprint
  {https://arxiv.org/abs/2105.04565} {arXiv:2105.04565 [hep-ph]} \BibitemShut
  {NoStop}%
\bibitem [{\citenamefont {Bennett}\ \emph {et~al.}(2006)\citenamefont {Bennett}
  \emph {et~al.}}]{Muong-2:2006rrc}%
  \BibitemOpen
  \bibfield  {author} {\bibinfo {author} {\bibfnamefont {G.~W.}\ \bibnamefont
  {Bennett}} \emph {et~al.} (\bibinfo {collaboration} {Muon g-2}),\ }\href
  {https://doi.org/10.1103/PhysRevD.73.072003} {\bibfield  {journal} {\bibinfo
  {journal} {Phys. Rev. D}\ }\textbf {\bibinfo {volume} {73}},\ \bibinfo
  {pages} {072003} (\bibinfo {year} {2006})},\ \Eprint
  {https://arxiv.org/abs/hep-ex/0602035} {arXiv:hep-ex/0602035} \BibitemShut
  {NoStop}%
\bibitem [{\citenamefont {Roberts}(2010)}]{Roberts:2010cj}%
  \BibitemOpen
  \bibfield  {author} {\bibinfo {author} {\bibfnamefont {B.~L.}\ \bibnamefont
  {Roberts}},\ }\href {https://doi.org/10.1088/1674-1137/34/6/021} {\bibfield
  {journal} {\bibinfo  {journal} {Chin. Phys. C}\ }\textbf {\bibinfo {volume}
  {34}},\ \bibinfo {pages} {741} (\bibinfo {year} {2010})},\ \Eprint
  {https://arxiv.org/abs/1001.2898} {arXiv:1001.2898 [hep-ex]} \BibitemShut
  {NoStop}%
\bibitem [{\citenamefont {Aoyama}\ \emph {et~al.}(2020)\citenamefont {Aoyama}
  \emph {et~al.}}]{Aoyama:2020ynm}%
  \BibitemOpen
  \bibfield  {author} {\bibinfo {author} {\bibfnamefont {T.}~\bibnamefont
  {Aoyama}} \emph {et~al.},\ }\href
  {https://doi.org/10.1016/j.physrep.2020.07.006} {\bibfield  {journal}
  {\bibinfo  {journal} {Phys. Rept.}\ }\textbf {\bibinfo {volume} {887}},\
  \bibinfo {pages} {1} (\bibinfo {year} {2020})},\ \Eprint
  {https://arxiv.org/abs/2006.04822} {arXiv:2006.04822 [hep-ph]} \BibitemShut
  {NoStop}%
\bibitem [{\citenamefont {Abi}\ \emph {et~al.}(2021)\citenamefont {Abi} \emph
  {et~al.}}]{Muong-2:2021ojo}%
  \BibitemOpen
  \bibfield  {author} {\bibinfo {author} {\bibfnamefont {B.}~\bibnamefont
  {Abi}} \emph {et~al.} (\bibinfo {collaboration} {Muon g-2}),\ }\href
  {https://doi.org/10.1103/PhysRevLett.126.141801} {\bibfield  {journal}
  {\bibinfo  {journal} {Phys. Rev. Lett.}\ }\textbf {\bibinfo {volume} {126}},\
  \bibinfo {pages} {141801} (\bibinfo {year} {2021})},\ \Eprint
  {https://arxiv.org/abs/2104.03281} {arXiv:2104.03281 [hep-ex]} \BibitemShut
  {NoStop}%
\bibitem [{\citenamefont {Feng}\ \emph {et~al.}(2016)\citenamefont {Feng},
  \citenamefont {Fornal}, \citenamefont {Galon}, \citenamefont {Gardner},
  \citenamefont {Smolinsky}, \citenamefont {Tait},\ and\ \citenamefont
  {Tanedo}}]{Feng:2016jff}%
  \BibitemOpen
  \bibfield  {author} {\bibinfo {author} {\bibfnamefont {J.~L.}\ \bibnamefont
  {Feng}}, \bibinfo {author} {\bibfnamefont {B.}~\bibnamefont {Fornal}},
  \bibinfo {author} {\bibfnamefont {I.}~\bibnamefont {Galon}}, \bibinfo
  {author} {\bibfnamefont {S.}~\bibnamefont {Gardner}}, \bibinfo {author}
  {\bibfnamefont {J.}~\bibnamefont {Smolinsky}}, \bibinfo {author}
  {\bibfnamefont {T.~M.~P.}\ \bibnamefont {Tait}},\ and\ \bibinfo {author}
  {\bibfnamefont {P.}~\bibnamefont {Tanedo}},\ }\href
  {https://doi.org/10.1103/PhysRevLett.117.071803} {\bibfield  {journal}
  {\bibinfo  {journal} {Phys. Rev. Lett.}\ }\textbf {\bibinfo {volume} {117}},\
  \bibinfo {pages} {071803} (\bibinfo {year} {2016})},\ \Eprint
  {https://arxiv.org/abs/1604.07411} {arXiv:1604.07411 [hep-ph]} \BibitemShut
  {NoStop}%
\bibitem [{\citenamefont {Feng}\ \emph {et~al.}(2017)\citenamefont {Feng},
  \citenamefont {Fornal}, \citenamefont {Galon}, \citenamefont {Gardner},
  \citenamefont {Smolinsky}, \citenamefont {Tait},\ and\ \citenamefont
  {Tanedo}}]{Feng:2016ysn}%
  \BibitemOpen
  \bibfield  {author} {\bibinfo {author} {\bibfnamefont {J.~L.}\ \bibnamefont
  {Feng}}, \bibinfo {author} {\bibfnamefont {B.}~\bibnamefont {Fornal}},
  \bibinfo {author} {\bibfnamefont {I.}~\bibnamefont {Galon}}, \bibinfo
  {author} {\bibfnamefont {S.}~\bibnamefont {Gardner}}, \bibinfo {author}
  {\bibfnamefont {J.}~\bibnamefont {Smolinsky}}, \bibinfo {author}
  {\bibfnamefont {T.~M.~P.}\ \bibnamefont {Tait}},\ and\ \bibinfo {author}
  {\bibfnamefont {P.}~\bibnamefont {Tanedo}},\ }\href
  {https://doi.org/10.1103/PhysRevD.95.035017} {\bibfield  {journal} {\bibinfo
  {journal} {Phys. Rev. D}\ }\textbf {\bibinfo {volume} {95}},\ \bibinfo
  {pages} {035017} (\bibinfo {year} {2017})},\ \Eprint
  {https://arxiv.org/abs/1608.03591} {arXiv:1608.03591 [hep-ph]} \BibitemShut
  {NoStop}%
\bibitem [{\citenamefont {Nelson}\ and\ \citenamefont
  {Scholtz}(2011)}]{Nelson:2011sf}%
  \BibitemOpen
  \bibfield  {author} {\bibinfo {author} {\bibfnamefont {A.~E.}\ \bibnamefont
  {Nelson}}\ and\ \bibinfo {author} {\bibfnamefont {J.}~\bibnamefont
  {Scholtz}},\ }\href {https://doi.org/10.1103/PhysRevD.84.103501} {\bibfield
  {journal} {\bibinfo  {journal} {Phys. Rev. D}\ }\textbf {\bibinfo {volume}
  {84}},\ \bibinfo {pages} {103501} (\bibinfo {year} {2011})},\ \Eprint
  {https://arxiv.org/abs/1105.2812} {arXiv:1105.2812 [hep-ph]} \BibitemShut
  {NoStop}%
\bibitem [{\citenamefont {Preskill}\ \emph {et~al.}(1983)\citenamefont
  {Preskill}, \citenamefont {Wise},\ and\ \citenamefont
  {Wilczek}}]{Preskill:1982cy}%
  \BibitemOpen
  \bibfield  {author} {\bibinfo {author} {\bibfnamefont {J.}~\bibnamefont
  {Preskill}}, \bibinfo {author} {\bibfnamefont {M.~B.}\ \bibnamefont {Wise}},\
  and\ \bibinfo {author} {\bibfnamefont {F.}~\bibnamefont {Wilczek}},\ }\href
  {https://doi.org/10.1016/0370-2693(83)90637-8} {\bibfield  {journal}
  {\bibinfo  {journal} {Phys. Lett. B}\ }\textbf {\bibinfo {volume} {120}},\
  \bibinfo {pages} {127} (\bibinfo {year} {1983})}\BibitemShut {NoStop}%
\bibitem [{\citenamefont {Abbott}\ and\ \citenamefont
  {Sikivie}(1983)}]{Abbott:1982af}%
  \BibitemOpen
  \bibfield  {author} {\bibinfo {author} {\bibfnamefont {L.~F.}\ \bibnamefont
  {Abbott}}\ and\ \bibinfo {author} {\bibfnamefont {P.}~\bibnamefont
  {Sikivie}},\ }\href {https://doi.org/10.1016/0370-2693(83)90638-X} {\bibfield
   {journal} {\bibinfo  {journal} {Phys. Lett. B}\ }\textbf {\bibinfo {volume}
  {120}},\ \bibinfo {pages} {133} (\bibinfo {year} {1983})}\BibitemShut
  {NoStop}%
\bibitem [{\citenamefont {Dine}\ and\ \citenamefont
  {Fischler}(1983)}]{Dine:1982ah}%
  \BibitemOpen
  \bibfield  {author} {\bibinfo {author} {\bibfnamefont {M.}~\bibnamefont
  {Dine}}\ and\ \bibinfo {author} {\bibfnamefont {W.}~\bibnamefont
  {Fischler}},\ }\href {https://doi.org/10.1016/0370-2693(83)90639-1}
  {\bibfield  {journal} {\bibinfo  {journal} {Phys. Lett. B}\ }\textbf
  {\bibinfo {volume} {120}},\ \bibinfo {pages} {137} (\bibinfo {year}
  {1983})}\BibitemShut {NoStop}%
\bibitem [{\citenamefont {Arias}\ \emph {et~al.}(2012)\citenamefont {Arias},
  \citenamefont {Cadamuro}, \citenamefont {Goodsell}, \citenamefont {Jaeckel},
  \citenamefont {Redondo},\ and\ \citenamefont {Ringwald}}]{Arias:2012az}%
  \BibitemOpen
  \bibfield  {author} {\bibinfo {author} {\bibfnamefont {P.}~\bibnamefont
  {Arias}}, \bibinfo {author} {\bibfnamefont {D.}~\bibnamefont {Cadamuro}},
  \bibinfo {author} {\bibfnamefont {M.}~\bibnamefont {Goodsell}}, \bibinfo
  {author} {\bibfnamefont {J.}~\bibnamefont {Jaeckel}}, \bibinfo {author}
  {\bibfnamefont {J.}~\bibnamefont {Redondo}},\ and\ \bibinfo {author}
  {\bibfnamefont {A.}~\bibnamefont {Ringwald}},\ }\href
  {https://doi.org/10.1088/1475-7516/2012/06/013} {\bibfield  {journal}
  {\bibinfo  {journal} {JCAP}\ }\textbf {\bibinfo {volume} {06}},\ \bibinfo
  {pages} {013}},\ \Eprint {https://arxiv.org/abs/1201.5902} {arXiv:1201.5902
  [hep-ph]} \BibitemShut {NoStop}%
\bibitem [{\citenamefont {Graham}\ \emph {et~al.}(2016)\citenamefont {Graham},
  \citenamefont {Mardon},\ and\ \citenamefont {Rajendran}}]{Graham:2015rva}%
  \BibitemOpen
  \bibfield  {author} {\bibinfo {author} {\bibfnamefont {P.~W.}\ \bibnamefont
  {Graham}}, \bibinfo {author} {\bibfnamefont {J.}~\bibnamefont {Mardon}},\
  and\ \bibinfo {author} {\bibfnamefont {S.}~\bibnamefont {Rajendran}},\ }\href
  {https://doi.org/10.1103/PhysRevD.93.103520} {\bibfield  {journal} {\bibinfo
  {journal} {Phys. Rev. D}\ }\textbf {\bibinfo {volume} {93}},\ \bibinfo
  {pages} {103520} (\bibinfo {year} {2016})},\ \Eprint
  {https://arxiv.org/abs/1504.02102} {arXiv:1504.02102 [hep-ph]} \BibitemShut
  {NoStop}%
\bibitem [{\citenamefont {Akrami}\ \emph {et~al.}(2020)\citenamefont {Akrami}
  \emph {et~al.}}]{Planck:2018jri}%
  \BibitemOpen
  \bibfield  {author} {\bibinfo {author} {\bibfnamefont {Y.}~\bibnamefont
  {Akrami}} \emph {et~al.} (\bibinfo {collaboration} {Planck}),\ }\href
  {https://doi.org/10.1051/0004-6361/201833887} {\bibfield  {journal} {\bibinfo
   {journal} {Astron. Astrophys.}\ }\textbf {\bibinfo {volume} {641}},\
  \bibinfo {pages} {A10} (\bibinfo {year} {2020})},\ \Eprint
  {https://arxiv.org/abs/1807.06211} {arXiv:1807.06211 [astro-ph.CO]}
  \BibitemShut {NoStop}%
\bibitem [{\citenamefont {Agrawal}\ \emph {et~al.}(2020)\citenamefont
  {Agrawal}, \citenamefont {Kitajima}, \citenamefont {Reece}, \citenamefont
  {Sekiguchi},\ and\ \citenamefont {Takahashi}}]{Agrawal:2018vin}%
  \BibitemOpen
  \bibfield  {author} {\bibinfo {author} {\bibfnamefont {P.}~\bibnamefont
  {Agrawal}}, \bibinfo {author} {\bibfnamefont {N.}~\bibnamefont {Kitajima}},
  \bibinfo {author} {\bibfnamefont {M.}~\bibnamefont {Reece}}, \bibinfo
  {author} {\bibfnamefont {T.}~\bibnamefont {Sekiguchi}},\ and\ \bibinfo
  {author} {\bibfnamefont {F.}~\bibnamefont {Takahashi}},\ }\href
  {https://doi.org/10.1016/j.physletb.2019.135136} {\bibfield  {journal}
  {\bibinfo  {journal} {Phys. Lett. B}\ }\textbf {\bibinfo {volume} {801}},\
  \bibinfo {pages} {135136} (\bibinfo {year} {2020})},\ \Eprint
  {https://arxiv.org/abs/1810.07188} {arXiv:1810.07188 [hep-ph]} \BibitemShut
  {NoStop}%
\bibitem [{\citenamefont {Dror}\ \emph {et~al.}(2019)\citenamefont {Dror},
  \citenamefont {Harigaya},\ and\ \citenamefont {Narayan}}]{Dror:2018pdh}%
  \BibitemOpen
  \bibfield  {author} {\bibinfo {author} {\bibfnamefont {J.~A.}\ \bibnamefont
  {Dror}}, \bibinfo {author} {\bibfnamefont {K.}~\bibnamefont {Harigaya}},\
  and\ \bibinfo {author} {\bibfnamefont {V.}~\bibnamefont {Narayan}},\ }\href
  {https://doi.org/10.1103/PhysRevD.99.035036} {\bibfield  {journal} {\bibinfo
  {journal} {Phys. Rev. D}\ }\textbf {\bibinfo {volume} {99}},\ \bibinfo
  {pages} {035036} (\bibinfo {year} {2019})},\ \Eprint
  {https://arxiv.org/abs/1810.07195} {arXiv:1810.07195 [hep-ph]} \BibitemShut
  {NoStop}%
\bibitem [{\citenamefont {Co}\ \emph {et~al.}(2019)\citenamefont {Co},
  \citenamefont {Pierce}, \citenamefont {Zhang},\ and\ \citenamefont
  {Zhao}}]{Co:2018lka}%
  \BibitemOpen
  \bibfield  {author} {\bibinfo {author} {\bibfnamefont {R.~T.}\ \bibnamefont
  {Co}}, \bibinfo {author} {\bibfnamefont {A.}~\bibnamefont {Pierce}}, \bibinfo
  {author} {\bibfnamefont {Z.}~\bibnamefont {Zhang}},\ and\ \bibinfo {author}
  {\bibfnamefont {Y.}~\bibnamefont {Zhao}},\ }\href
  {https://doi.org/10.1103/PhysRevD.99.075002} {\bibfield  {journal} {\bibinfo
  {journal} {Phys. Rev. D}\ }\textbf {\bibinfo {volume} {99}},\ \bibinfo
  {pages} {075002} (\bibinfo {year} {2019})},\ \Eprint
  {https://arxiv.org/abs/1810.07196} {arXiv:1810.07196 [hep-ph]} \BibitemShut
  {NoStop}%
\bibitem [{\citenamefont {Bastero-Gil}\ \emph {et~al.}(2019)\citenamefont
  {Bastero-Gil}, \citenamefont {Santiago}, \citenamefont {Ubaldi},\ and\
  \citenamefont {Vega-Morales}}]{Bastero-Gil:2018uel}%
  \BibitemOpen
  \bibfield  {author} {\bibinfo {author} {\bibfnamefont {M.}~\bibnamefont
  {Bastero-Gil}}, \bibinfo {author} {\bibfnamefont {J.}~\bibnamefont
  {Santiago}}, \bibinfo {author} {\bibfnamefont {L.}~\bibnamefont {Ubaldi}},\
  and\ \bibinfo {author} {\bibfnamefont {R.}~\bibnamefont {Vega-Morales}},\
  }\href {https://doi.org/10.1088/1475-7516/2019/04/015} {\bibfield  {journal}
  {\bibinfo  {journal} {JCAP}\ }\textbf {\bibinfo {volume} {04}},\ \bibinfo
  {pages} {015}},\ \Eprint {https://arxiv.org/abs/1810.07208} {arXiv:1810.07208
  [hep-ph]} \BibitemShut {NoStop}%
\bibitem [{\citenamefont {Co}\ \emph {et~al.}(2021)\citenamefont {Co},
  \citenamefont {Harigaya},\ and\ \citenamefont {Pierce}}]{Co:2021rhi}%
  \BibitemOpen
  \bibfield  {author} {\bibinfo {author} {\bibfnamefont {R.~T.}\ \bibnamefont
  {Co}}, \bibinfo {author} {\bibfnamefont {K.}~\bibnamefont {Harigaya}},\ and\
  \bibinfo {author} {\bibfnamefont {A.}~\bibnamefont {Pierce}},\ }\href
  {https://doi.org/10.1007/JHEP12(2021)099} {\bibfield  {journal} {\bibinfo
  {journal} {JHEP}\ }\textbf {\bibinfo {volume} {12}},\ \bibinfo {pages}
  {099}},\ \Eprint {https://arxiv.org/abs/2104.02077} {arXiv:2104.02077
  [hep-ph]} \BibitemShut {NoStop}%
\bibitem [{\citenamefont {Long}\ and\ \citenamefont
  {Wang}(2019)}]{Long:2019lwl}%
  \BibitemOpen
  \bibfield  {author} {\bibinfo {author} {\bibfnamefont {A.~J.}\ \bibnamefont
  {Long}}\ and\ \bibinfo {author} {\bibfnamefont {L.-T.}\ \bibnamefont
  {Wang}},\ }\href {https://doi.org/10.1103/PhysRevD.99.063529} {\bibfield
  {journal} {\bibinfo  {journal} {Phys. Rev. D}\ }\textbf {\bibinfo {volume}
  {99}},\ \bibinfo {pages} {063529} (\bibinfo {year} {2019})},\ \Eprint
  {https://arxiv.org/abs/1901.03312} {arXiv:1901.03312 [hep-ph]} \BibitemShut
  {NoStop}%
\bibitem [{\citenamefont {Silin}(1960)}]{Silin:1960pya}%
  \BibitemOpen
  \bibfield  {author} {\bibinfo {author} {\bibfnamefont {V.~P.}\ \bibnamefont
  {Silin}},\ }\href@noop {} {\bibfield  {journal} {\bibinfo  {journal} {Sov.
  Phys. JETP}\ }\textbf {\bibinfo {volume} {11}},\ \bibinfo {pages} {1136}
  (\bibinfo {year} {1960})}\BibitemShut {NoStop}%
\bibitem [{\citenamefont {{Jancovici}}(1962)}]{1962NCim...25..428J}%
  \BibitemOpen
  \bibfield  {author} {\bibinfo {author} {\bibfnamefont {B.}~\bibnamefont
  {{Jancovici}}},\ }\href {https://doi.org/10.1007/BF02731458} {\bibfield
  {journal} {\bibinfo  {journal} {Il Nuovo Cimento}\ }\textbf {\bibinfo
  {volume} {25}},\ \bibinfo {pages} {428} (\bibinfo {year} {1962})}\BibitemShut
  {NoStop}%
\bibitem [{\citenamefont {Beaudet}\ \emph {et~al.}(1967)\citenamefont
  {Beaudet}, \citenamefont {Petrosian},\ and\ \citenamefont
  {Salpeter}}]{Beaudet:1967zz}%
  \BibitemOpen
  \bibfield  {author} {\bibinfo {author} {\bibfnamefont {G.}~\bibnamefont
  {Beaudet}}, \bibinfo {author} {\bibfnamefont {V.}~\bibnamefont {Petrosian}},\
  and\ \bibinfo {author} {\bibfnamefont {E.~E.}\ \bibnamefont {Salpeter}},\
  }\href {https://doi.org/10.1086/149398} {\bibfield  {journal} {\bibinfo
  {journal} {Astrophys. J.}\ }\textbf {\bibinfo {volume} {150}},\ \bibinfo
  {pages} {979} (\bibinfo {year} {1967})}\BibitemShut {NoStop}%
\bibitem [{\citenamefont {Klimov}(1982)}]{Klimov:1982bv}%
  \BibitemOpen
  \bibfield  {author} {\bibinfo {author} {\bibfnamefont {V.~V.}\ \bibnamefont
  {Klimov}},\ }\href@noop {} {\bibfield  {journal} {\bibinfo  {journal} {Sov.
  Phys. JETP}\ }\textbf {\bibinfo {volume} {55}},\ \bibinfo {pages} {199}
  (\bibinfo {year} {1982})}\BibitemShut {NoStop}%
\bibitem [{\citenamefont {Weldon}(1982)}]{Weldon:1982aq}%
  \BibitemOpen
  \bibfield  {author} {\bibinfo {author} {\bibfnamefont {H.~A.}\ \bibnamefont
  {Weldon}},\ }\href {https://doi.org/10.1103/PhysRevD.26.1394} {\bibfield
  {journal} {\bibinfo  {journal} {Phys. Rev. D}\ }\textbf {\bibinfo {volume}
  {26}},\ \bibinfo {pages} {1394} (\bibinfo {year} {1982})}\BibitemShut
  {NoStop}%
\bibitem [{\citenamefont {Braaten}\ and\ \citenamefont
  {Segel}(1993)}]{Braaten:1993jw}%
  \BibitemOpen
  \bibfield  {author} {\bibinfo {author} {\bibfnamefont {E.}~\bibnamefont
  {Braaten}}\ and\ \bibinfo {author} {\bibfnamefont {D.}~\bibnamefont
  {Segel}},\ }\href {https://doi.org/10.1103/PhysRevD.48.1478} {\bibfield
  {journal} {\bibinfo  {journal} {Phys. Rev. D}\ }\textbf {\bibinfo {volume}
  {48}},\ \bibinfo {pages} {1478} (\bibinfo {year} {1993})},\ \Eprint
  {https://arxiv.org/abs/hep-ph/9302213} {arXiv:hep-ph/9302213} \BibitemShut
  {NoStop}%
\bibitem [{\citenamefont {Choi}\ \emph
  {et~al.}(2020{\natexlab{a}})\citenamefont {Choi}, \citenamefont {Chun},\ and\
  \citenamefont {Kim}}]{Choi:2019zxy}%
  \BibitemOpen
  \bibfield  {author} {\bibinfo {author} {\bibfnamefont {K.-Y.}\ \bibnamefont
  {Choi}}, \bibinfo {author} {\bibfnamefont {E.~J.}\ \bibnamefont {Chun}},\
  and\ \bibinfo {author} {\bibfnamefont {J.}~\bibnamefont {Kim}},\ }\href
  {https://doi.org/10.1016/j.dark.2020.100606} {\bibfield  {journal} {\bibinfo
  {journal} {Phys. Dark Univ.}\ }\textbf {\bibinfo {volume} {30}},\ \bibinfo
  {pages} {100606} (\bibinfo {year} {2020}{\natexlab{a}})},\ \Eprint
  {https://arxiv.org/abs/1909.10478} {arXiv:1909.10478 [hep-ph]} \BibitemShut
  {NoStop}%
\bibitem [{\citenamefont {Choi}\ \emph
  {et~al.}(2020{\natexlab{b}})\citenamefont {Choi}, \citenamefont {Chun},\ and\
  \citenamefont {Kim}}]{Choi:2020ydp}%
  \BibitemOpen
  \bibfield  {author} {\bibinfo {author} {\bibfnamefont {K.-Y.}\ \bibnamefont
  {Choi}}, \bibinfo {author} {\bibfnamefont {E.~J.}\ \bibnamefont {Chun}},\
  and\ \bibinfo {author} {\bibfnamefont {J.}~\bibnamefont {Kim}},\ }\href@noop
  {} {\  (\bibinfo {year} {2020}{\natexlab{b}})},\ \Eprint
  {https://arxiv.org/abs/2012.09474} {arXiv:2012.09474 [hep-ph]} \BibitemShut
  {NoStop}%
\bibitem [{Note1()}]{Note1}%
  \BibitemOpen
  \bibinfo {note} {Recently, the impact of the small scale structure in vector
  dark matter on its polarizations has been discussed in Ref.~\cite
  {Amin:2022pzv}.}\BibitemShut {Stop}%
\bibitem [{\citenamefont {Arkani-Hamed}\ \emph {et~al.}(2009)\citenamefont
  {Arkani-Hamed}, \citenamefont {Finkbeiner}, \citenamefont {Slatyer},\ and\
  \citenamefont {Weiner}}]{Arkani-Hamed:2008hhe}%
  \BibitemOpen
  \bibfield  {author} {\bibinfo {author} {\bibfnamefont {N.}~\bibnamefont
  {Arkani-Hamed}}, \bibinfo {author} {\bibfnamefont {D.~P.}\ \bibnamefont
  {Finkbeiner}}, \bibinfo {author} {\bibfnamefont {T.~R.}\ \bibnamefont
  {Slatyer}},\ and\ \bibinfo {author} {\bibfnamefont {N.}~\bibnamefont
  {Weiner}},\ }\href {https://doi.org/10.1103/PhysRevD.79.015014} {\bibfield
  {journal} {\bibinfo  {journal} {Phys. Rev. D}\ }\textbf {\bibinfo {volume}
  {79}},\ \bibinfo {pages} {015014} (\bibinfo {year} {2009})},\ \Eprint
  {https://arxiv.org/abs/0810.0713} {arXiv:0810.0713 [hep-ph]} \BibitemShut
  {NoStop}%
\bibitem [{Note2()}]{Note2}%
  \BibitemOpen
  \bibinfo {note} {There is also the correction to the 2-point function from
  the same diagram of Fig.~\ref {fig:SelfE} with the quantum loop (i.e., the
  quantum flunctuated propagator with a loop momentum). Due to the loop
  suppression, we ignore this radiatively induced contribution.}\BibitemShut
  {Stop}%
\bibitem [{\citenamefont {Ward}(1950)}]{Ward:1950xp}%
  \BibitemOpen
  \bibfield  {author} {\bibinfo {author} {\bibfnamefont {J.~C.}\ \bibnamefont
  {Ward}},\ }\href {https://doi.org/10.1103/PhysRev.78.182} {\bibfield
  {journal} {\bibinfo  {journal} {Phys. Rev.}\ }\textbf {\bibinfo {volume}
  {78}},\ \bibinfo {pages} {182} (\bibinfo {year} {1950})}\BibitemShut
  {NoStop}%
\bibitem [{\citenamefont {Takahashi}(1957)}]{Takahashi:1957xn}%
  \BibitemOpen
  \bibfield  {author} {\bibinfo {author} {\bibfnamefont {Y.}~\bibnamefont
  {Takahashi}},\ }\href {https://doi.org/10.1007/BF02832514} {\bibfield
  {journal} {\bibinfo  {journal} {Nuovo Cim.}\ }\textbf {\bibinfo {volume}
  {6}},\ \bibinfo {pages} {371} (\bibinfo {year} {1957})}\BibitemShut {NoStop}%
\bibitem [{\citenamefont {Penning}(1936)}]{PENNING1936873}%
  \BibitemOpen
  \bibfield  {author} {\bibinfo {author} {\bibfnamefont {F.}~\bibnamefont
  {Penning}},\ }\href
  {https://doi.org/https://doi.org/10.1016/S0031-8914(36)80313-9} {\bibfield
  {journal} {\bibinfo  {journal} {Physica}\ }\textbf {\bibinfo {volume} {3}},\
  \bibinfo {pages} {873} (\bibinfo {year} {1936})}\BibitemShut {NoStop}%
\bibitem [{\citenamefont {Pierce}(1949)}]{PhysRev.76.565.2}%
  \BibitemOpen
  \bibfield  {author} {\bibinfo {author} {\bibfnamefont {J.~R.}\ \bibnamefont
  {Pierce}},\ }\href {https://doi.org/10.1103/PhysRev.76.565.2} {\bibfield
  {journal} {\bibinfo  {journal} {Phys. Rev.}\ }\textbf {\bibinfo {volume}
  {76}},\ \bibinfo {pages} {565} (\bibinfo {year} {1949})}\BibitemShut
  {NoStop}%
\bibitem [{\citenamefont {Brown}\ and\ \citenamefont
  {Gabrielse}(1986)}]{Brown:1985rh}%
  \BibitemOpen
  \bibfield  {author} {\bibinfo {author} {\bibfnamefont {L.~S.}\ \bibnamefont
  {Brown}}\ and\ \bibinfo {author} {\bibfnamefont {G.}~\bibnamefont
  {Gabrielse}},\ }\href {https://doi.org/10.1103/RevModPhys.58.233} {\bibfield
  {journal} {\bibinfo  {journal} {Rev. Mod. Phys.}\ }\textbf {\bibinfo {volume}
  {58}},\ \bibinfo {pages} {233} (\bibinfo {year} {1986})}\BibitemShut
  {NoStop}%
\bibitem [{\citenamefont {Kerr}\ and\ \citenamefont
  {Lynden-Bell}(1986)}]{Kerr:1986hz}%
  \BibitemOpen
  \bibfield  {author} {\bibinfo {author} {\bibfnamefont {F.~J.}\ \bibnamefont
  {Kerr}}\ and\ \bibinfo {author} {\bibfnamefont {D.}~\bibnamefont
  {Lynden-Bell}},\ }\href@noop {} {\bibfield  {journal} {\bibinfo  {journal}
  {Mon. Not. Roy. Astron. Soc.}\ }\textbf {\bibinfo {volume} {221}},\ \bibinfo
  {pages} {1023} (\bibinfo {year} {1986})}\BibitemShut {NoStop}%
\bibitem [{\citenamefont {Reid}\ \emph {et~al.}(2009)\citenamefont {Reid} \emph
  {et~al.}}]{Reid:2009nj}%
  \BibitemOpen
  \bibfield  {author} {\bibinfo {author} {\bibfnamefont {M.~J.}\ \bibnamefont
  {Reid}} \emph {et~al.},\ }\href {https://doi.org/10.1088/0004-637X/700/1/137}
  {\bibfield  {journal} {\bibinfo  {journal} {Astrophys. J.}\ }\textbf
  {\bibinfo {volume} {700}},\ \bibinfo {pages} {137} (\bibinfo {year}
  {2009})},\ \Eprint {https://arxiv.org/abs/0902.3913} {arXiv:0902.3913
  [astro-ph.GA]} \BibitemShut {NoStop}%
\bibitem [{\citenamefont {McMillan}\ and\ \citenamefont
  {Binney}(2010)}]{McMillan:2009yr}%
  \BibitemOpen
  \bibfield  {author} {\bibinfo {author} {\bibfnamefont {P.~J.}\ \bibnamefont
  {McMillan}}\ and\ \bibinfo {author} {\bibfnamefont {J.~J.}\ \bibnamefont
  {Binney}},\ }\href {https://doi.org/10.1111/j.1365-2966.2009.15932.x}
  {\bibfield  {journal} {\bibinfo  {journal} {Mon. Not. Roy. Astron. Soc.}\
  }\textbf {\bibinfo {volume} {402}},\ \bibinfo {pages} {934} (\bibinfo {year}
  {2010})},\ \Eprint {https://arxiv.org/abs/0907.4685} {arXiv:0907.4685
  [astro-ph.GA]} \BibitemShut {NoStop}%
\bibitem [{\citenamefont {Bovy}\ \emph {et~al.}(2009)\citenamefont {Bovy},
  \citenamefont {Hogg},\ and\ \citenamefont {Rix}}]{Bovy:2009dr}%
  \BibitemOpen
  \bibfield  {author} {\bibinfo {author} {\bibfnamefont {J.}~\bibnamefont
  {Bovy}}, \bibinfo {author} {\bibfnamefont {D.~W.}\ \bibnamefont {Hogg}},\
  and\ \bibinfo {author} {\bibfnamefont {H.-W.}\ \bibnamefont {Rix}},\ }\href
  {https://doi.org/10.1088/0004-637X/704/2/1704} {\bibfield  {journal}
  {\bibinfo  {journal} {Astrophys. J.}\ }\textbf {\bibinfo {volume} {704}},\
  \bibinfo {pages} {1704} (\bibinfo {year} {2009})},\ \Eprint
  {https://arxiv.org/abs/0907.5423} {arXiv:0907.5423 [astro-ph.GA]}
  \BibitemShut {NoStop}%
\bibitem [{\citenamefont {Freese}\ \emph {et~al.}(2013)\citenamefont {Freese},
  \citenamefont {Lisanti},\ and\ \citenamefont {Savage}}]{Freese:2012xd}%
  \BibitemOpen
  \bibfield  {author} {\bibinfo {author} {\bibfnamefont {K.}~\bibnamefont
  {Freese}}, \bibinfo {author} {\bibfnamefont {M.}~\bibnamefont {Lisanti}},\
  and\ \bibinfo {author} {\bibfnamefont {C.}~\bibnamefont {Savage}},\ }\href
  {https://doi.org/10.1103/RevModPhys.85.1561} {\bibfield  {journal} {\bibinfo
  {journal} {Rev. Mod. Phys.}\ }\textbf {\bibinfo {volume} {85}},\ \bibinfo
  {pages} {1561} (\bibinfo {year} {2013})},\ \Eprint
  {https://arxiv.org/abs/1209.3339} {arXiv:1209.3339 [astro-ph.CO]}
  \BibitemShut {NoStop}%
\bibitem [{\citenamefont {Zyla}\ \emph {et~al.}(2020)\citenamefont {Zyla} \emph
  {et~al.}}]{Zyla:2020zbs}%
  \BibitemOpen
  \bibfield  {author} {\bibinfo {author} {\bibfnamefont {P.}~\bibnamefont
  {Zyla}} \emph {et~al.} (\bibinfo {collaboration} {Particle Data Group}),\
  }\href {https://doi.org/10.1093/ptep/ptaa104} {\bibfield  {journal} {\bibinfo
   {journal} {PTEP}\ }\textbf {\bibinfo {volume} {2020}},\ \bibinfo {pages}
  {083C01} (\bibinfo {year} {2020})},\ \bibinfo {note} {and 2021
  update}\BibitemShut {NoStop}%
\bibitem [{\citenamefont {Benito}\ \emph {et~al.}(2019)\citenamefont {Benito},
  \citenamefont {Cuoco},\ and\ \citenamefont {Iocco}}]{Benito:2019ngh}%
  \BibitemOpen
  \bibfield  {author} {\bibinfo {author} {\bibfnamefont {M.}~\bibnamefont
  {Benito}}, \bibinfo {author} {\bibfnamefont {A.}~\bibnamefont {Cuoco}},\ and\
  \bibinfo {author} {\bibfnamefont {F.}~\bibnamefont {Iocco}},\ }\href
  {https://doi.org/10.1088/1475-7516/2019/03/033} {\bibfield  {journal}
  {\bibinfo  {journal} {JCAP}\ }\textbf {\bibinfo {volume} {03}},\ \bibinfo
  {pages} {033}},\ \Eprint {https://arxiv.org/abs/1901.02460} {arXiv:1901.02460
  [astro-ph.GA]} \BibitemShut {NoStop}%
\bibitem [{\citenamefont {Mather}\ \emph {et~al.}(1999)\citenamefont {Mather},
  \citenamefont {Fixsen}, \citenamefont {Shafer}, \citenamefont {Mosier},\ and\
  \citenamefont {Wilkinson}}]{Mather:1998gm}%
  \BibitemOpen
  \bibfield  {author} {\bibinfo {author} {\bibfnamefont {J.~C.}\ \bibnamefont
  {Mather}}, \bibinfo {author} {\bibfnamefont {D.~J.}\ \bibnamefont {Fixsen}},
  \bibinfo {author} {\bibfnamefont {R.~A.}\ \bibnamefont {Shafer}}, \bibinfo
  {author} {\bibfnamefont {C.}~\bibnamefont {Mosier}},\ and\ \bibinfo {author}
  {\bibfnamefont {D.~T.}\ \bibnamefont {Wilkinson}},\ }\href
  {https://doi.org/10.1086/306805} {\bibfield  {journal} {\bibinfo  {journal}
  {Astrophys. J.}\ }\textbf {\bibinfo {volume} {512}},\ \bibinfo {pages} {511}
  (\bibinfo {year} {1999})},\ \Eprint {https://arxiv.org/abs/astro-ph/9810373}
  {arXiv:astro-ph/9810373} \BibitemShut {NoStop}%
\bibitem [{\citenamefont {Fixsen}(2009)}]{Fixsen:2009ug}%
  \BibitemOpen
  \bibfield  {author} {\bibinfo {author} {\bibfnamefont {D.~J.}\ \bibnamefont
  {Fixsen}},\ }\href {https://doi.org/10.1088/0004-637X/707/2/916} {\bibfield
  {journal} {\bibinfo  {journal} {Astrophys. J.}\ }\textbf {\bibinfo {volume}
  {707}},\ \bibinfo {pages} {916} (\bibinfo {year} {2009})},\ \Eprint
  {https://arxiv.org/abs/0911.1955} {arXiv:0911.1955 [astro-ph.CO]}
  \BibitemShut {NoStop}%
\bibitem [{\citenamefont {Noterdaeme}\ \emph {et~al.}(2011)\citenamefont
  {Noterdaeme}, \citenamefont {Petitjean}, \citenamefont {Srianand},
  \citenamefont {Ledoux},\ and\ \citenamefont {Lopez}}]{Noterdaeme:2010tm}%
  \BibitemOpen
  \bibfield  {author} {\bibinfo {author} {\bibfnamefont {P.}~\bibnamefont
  {Noterdaeme}}, \bibinfo {author} {\bibfnamefont {P.}~\bibnamefont
  {Petitjean}}, \bibinfo {author} {\bibfnamefont {R.}~\bibnamefont {Srianand}},
  \bibinfo {author} {\bibfnamefont {C.}~\bibnamefont {Ledoux}},\ and\ \bibinfo
  {author} {\bibfnamefont {S.}~\bibnamefont {Lopez}},\ }\href
  {https://doi.org/10.1051/0004-6361/201016140} {\bibfield  {journal} {\bibinfo
   {journal} {Astron. Astrophys.}\ }\textbf {\bibinfo {volume} {526}},\
  \bibinfo {pages} {L7} (\bibinfo {year} {2011})},\ \Eprint
  {https://arxiv.org/abs/1012.3164} {arXiv:1012.3164 [astro-ph.CO]}
  \BibitemShut {NoStop}%
\bibitem [{\citenamefont {de~Salas}\ \emph {et~al.}(2018)\citenamefont
  {de~Salas}, \citenamefont {Forero}, \citenamefont {Ternes}, \citenamefont
  {Tortola},\ and\ \citenamefont {Valle}}]{deSalas:2017kay}%
  \BibitemOpen
  \bibfield  {author} {\bibinfo {author} {\bibfnamefont {P.~F.}\ \bibnamefont
  {de~Salas}}, \bibinfo {author} {\bibfnamefont {D.~V.}\ \bibnamefont
  {Forero}}, \bibinfo {author} {\bibfnamefont {C.~A.}\ \bibnamefont {Ternes}},
  \bibinfo {author} {\bibfnamefont {M.}~\bibnamefont {Tortola}},\ and\ \bibinfo
  {author} {\bibfnamefont {J.~W.~F.}\ \bibnamefont {Valle}},\ }\href
  {https://doi.org/10.1016/j.physletb.2018.06.019} {\bibfield  {journal}
  {\bibinfo  {journal} {Phys. Lett. B}\ }\textbf {\bibinfo {volume} {782}},\
  \bibinfo {pages} {633} (\bibinfo {year} {2018})},\ \Eprint
  {https://arxiv.org/abs/1708.01186} {arXiv:1708.01186 [hep-ph]} \BibitemShut
  {NoStop}%
\bibitem [{\citenamefont {Capozzi}\ \emph {et~al.}(2018)\citenamefont
  {Capozzi}, \citenamefont {Lisi}, \citenamefont {Marrone},\ and\ \citenamefont
  {Palazzo}}]{Capozzi:2018ubv}%
  \BibitemOpen
  \bibfield  {author} {\bibinfo {author} {\bibfnamefont {F.}~\bibnamefont
  {Capozzi}}, \bibinfo {author} {\bibfnamefont {E.}~\bibnamefont {Lisi}},
  \bibinfo {author} {\bibfnamefont {A.}~\bibnamefont {Marrone}},\ and\ \bibinfo
  {author} {\bibfnamefont {A.}~\bibnamefont {Palazzo}},\ }\href
  {https://doi.org/10.1016/j.ppnp.2018.05.005} {\bibfield  {journal} {\bibinfo
  {journal} {Prog. Part. Nucl. Phys.}\ }\textbf {\bibinfo {volume} {102}},\
  \bibinfo {pages} {48} (\bibinfo {year} {2018})},\ \Eprint
  {https://arxiv.org/abs/1804.09678} {arXiv:1804.09678 [hep-ph]} \BibitemShut
  {NoStop}%
\bibitem [{\citenamefont {Esteban}\ \emph {et~al.}(2019)\citenamefont
  {Esteban}, \citenamefont {Gonzalez-Garcia}, \citenamefont
  {Hernandez-Cabezudo}, \citenamefont {Maltoni},\ and\ \citenamefont
  {Schwetz}}]{Esteban:2018azc}%
  \BibitemOpen
  \bibfield  {author} {\bibinfo {author} {\bibfnamefont {I.}~\bibnamefont
  {Esteban}}, \bibinfo {author} {\bibfnamefont {M.~C.}\ \bibnamefont
  {Gonzalez-Garcia}}, \bibinfo {author} {\bibfnamefont {A.}~\bibnamefont
  {Hernandez-Cabezudo}}, \bibinfo {author} {\bibfnamefont {M.}~\bibnamefont
  {Maltoni}},\ and\ \bibinfo {author} {\bibfnamefont {T.}~\bibnamefont
  {Schwetz}},\ }\href {https://doi.org/10.1007/JHEP01(2019)106} {\bibfield
  {journal} {\bibinfo  {journal} {JHEP}\ }\textbf {\bibinfo {volume} {01}},\
  \bibinfo {pages} {106}},\ \Eprint {https://arxiv.org/abs/1811.05487}
  {arXiv:1811.05487 [hep-ph]} \BibitemShut {NoStop}%
\bibitem [{\citenamefont {Esteban}\ \emph {et~al.}(2020)\citenamefont
  {Esteban}, \citenamefont {Gonzalez-Garcia}, \citenamefont {Maltoni},
  \citenamefont {Schwetz},\ and\ \citenamefont {Zhou}}]{Esteban:2020cvm}%
  \BibitemOpen
  \bibfield  {author} {\bibinfo {author} {\bibfnamefont {I.}~\bibnamefont
  {Esteban}}, \bibinfo {author} {\bibfnamefont {M.~C.}\ \bibnamefont
  {Gonzalez-Garcia}}, \bibinfo {author} {\bibfnamefont {M.}~\bibnamefont
  {Maltoni}}, \bibinfo {author} {\bibfnamefont {T.}~\bibnamefont {Schwetz}},\
  and\ \bibinfo {author} {\bibfnamefont {A.}~\bibnamefont {Zhou}},\ }\href
  {https://doi.org/10.1007/JHEP09(2020)178} {\bibfield  {journal} {\bibinfo
  {journal} {JHEP}\ }\textbf {\bibinfo {volume} {09}},\ \bibinfo {pages}
  {178}},\ \Eprint {https://arxiv.org/abs/2007.14792} {arXiv:2007.14792
  [hep-ph]} \BibitemShut {NoStop}%
\bibitem [{\citenamefont {Aker}\ \emph {et~al.}(2019)\citenamefont {Aker} \emph
  {et~al.}}]{KATRIN:2019yun}%
  \BibitemOpen
  \bibfield  {author} {\bibinfo {author} {\bibfnamefont {M.}~\bibnamefont
  {Aker}} \emph {et~al.} (\bibinfo {collaboration} {KATRIN}),\ }\href
  {https://doi.org/10.1103/PhysRevLett.123.221802} {\bibfield  {journal}
  {\bibinfo  {journal} {Phys. Rev. Lett.}\ }\textbf {\bibinfo {volume} {123}},\
  \bibinfo {pages} {221802} (\bibinfo {year} {2019})},\ \Eprint
  {https://arxiv.org/abs/1909.06048} {arXiv:1909.06048 [hep-ex]} \BibitemShut
  {NoStop}%
\bibitem [{\citenamefont {Aker}\ \emph {et~al.}(2022)\citenamefont {Aker} \emph
  {et~al.}}]{KATRIN:2021uub}%
  \BibitemOpen
  \bibfield  {author} {\bibinfo {author} {\bibfnamefont {M.}~\bibnamefont
  {Aker}} \emph {et~al.} (\bibinfo {collaboration} {KATRIN}),\ }\href
  {https://doi.org/10.1038/s41567-021-01463-1} {\bibfield  {journal} {\bibinfo
  {journal} {Nature Phys.}\ }\textbf {\bibinfo {volume} {18}},\ \bibinfo
  {pages} {160} (\bibinfo {year} {2022})},\ \Eprint
  {https://arxiv.org/abs/2105.08533} {arXiv:2105.08533 [hep-ex]} \BibitemShut
  {NoStop}%
\bibitem [{\citenamefont {Aghanim}\ \emph {et~al.}(2020)\citenamefont {Aghanim}
  \emph {et~al.}}]{Planck:2018vyg}%
  \BibitemOpen
  \bibfield  {author} {\bibinfo {author} {\bibfnamefont {N.}~\bibnamefont
  {Aghanim}} \emph {et~al.} (\bibinfo {collaboration} {Planck}),\ }\href
  {https://doi.org/10.1051/0004-6361/201833910} {\bibfield  {journal} {\bibinfo
   {journal} {Astron. Astrophys.}\ }\textbf {\bibinfo {volume} {641}},\
  \bibinfo {pages} {A6} (\bibinfo {year} {2020})},\ \bibinfo {note} {[Erratum:
  Astron.Astrophys. 652, C4 (2021)]},\ \Eprint
  {https://arxiv.org/abs/1807.06209} {arXiv:1807.06209 [astro-ph.CO]}
  \BibitemShut {NoStop}%
\bibitem [{\citenamefont {Wolfenstein}(1978)}]{Wolfenstein:1977ue}%
  \BibitemOpen
  \bibfield  {author} {\bibinfo {author} {\bibfnamefont {L.}~\bibnamefont
  {Wolfenstein}},\ }\href {https://doi.org/10.1103/PhysRevD.17.2369} {\bibfield
   {journal} {\bibinfo  {journal} {Phys. Rev. D}\ }\textbf {\bibinfo {volume}
  {17}},\ \bibinfo {pages} {2369} (\bibinfo {year} {1978})}\BibitemShut
  {NoStop}%
\bibitem [{\citenamefont {Mikheyev}\ and\ \citenamefont
  {Smirnov}(1985)}]{Mikheyev:1985zog}%
  \BibitemOpen
  \bibfield  {author} {\bibinfo {author} {\bibfnamefont {S.~P.}\ \bibnamefont
  {Mikheyev}}\ and\ \bibinfo {author} {\bibfnamefont {A.~Y.}\ \bibnamefont
  {Smirnov}},\ }\href@noop {} {\bibfield  {journal} {\bibinfo  {journal} {Sov.
  J. Nucl. Phys.}\ }\textbf {\bibinfo {volume} {42}},\ \bibinfo {pages} {913}
  (\bibinfo {year} {1985})}\BibitemShut {NoStop}%
\bibitem [{\citenamefont {Coloma}\ \emph {et~al.}(2021)\citenamefont {Coloma},
  \citenamefont {Gonzalez-Garcia},\ and\ \citenamefont
  {Maltoni}}]{Coloma:2020gfv}%
  \BibitemOpen
  \bibfield  {author} {\bibinfo {author} {\bibfnamefont {P.}~\bibnamefont
  {Coloma}}, \bibinfo {author} {\bibfnamefont {M.~C.}\ \bibnamefont
  {Gonzalez-Garcia}},\ and\ \bibinfo {author} {\bibfnamefont {M.}~\bibnamefont
  {Maltoni}},\ }\href {https://doi.org/10.1007/JHEP01(2021)114} {\bibfield
  {journal} {\bibinfo  {journal} {JHEP}\ }\textbf {\bibinfo {volume} {01}},\
  \bibinfo {pages} {114}},\ \Eprint {https://arxiv.org/abs/2009.14220}
  {arXiv:2009.14220 [hep-ph]} \BibitemShut {NoStop}%
\bibitem [{\citenamefont {Bahcall}\ \emph {et~al.}(2001)\citenamefont
  {Bahcall}, \citenamefont {Pinsonneault},\ and\ \citenamefont
  {Basu}}]{Bahcall:2000nu}%
  \BibitemOpen
  \bibfield  {author} {\bibinfo {author} {\bibfnamefont {J.~N.}\ \bibnamefont
  {Bahcall}}, \bibinfo {author} {\bibfnamefont {M.~H.}\ \bibnamefont
  {Pinsonneault}},\ and\ \bibinfo {author} {\bibfnamefont {S.}~\bibnamefont
  {Basu}},\ }\href {https://doi.org/10.1086/321493} {\bibfield  {journal}
  {\bibinfo  {journal} {Astrophys. J.}\ }\textbf {\bibinfo {volume} {555}},\
  \bibinfo {pages} {990} (\bibinfo {year} {2001})},\ \Eprint
  {https://arxiv.org/abs/astro-ph/0010346} {arXiv:astro-ph/0010346}
  \BibitemShut {NoStop}%
\bibitem [{\citenamefont {Serenelli}\ \emph {et~al.}(2011)\citenamefont
  {Serenelli}, \citenamefont {Haxton},\ and\ \citenamefont
  {Pena-Garay}}]{Serenelli:2011py}%
  \BibitemOpen
  \bibfield  {author} {\bibinfo {author} {\bibfnamefont {A.~M.}\ \bibnamefont
  {Serenelli}}, \bibinfo {author} {\bibfnamefont {W.~C.}\ \bibnamefont
  {Haxton}},\ and\ \bibinfo {author} {\bibfnamefont {C.}~\bibnamefont
  {Pena-Garay}},\ }\href {https://doi.org/10.1088/0004-637X/743/1/24}
  {\bibfield  {journal} {\bibinfo  {journal} {Astrophys. J.}\ }\textbf
  {\bibinfo {volume} {743}},\ \bibinfo {pages} {24} (\bibinfo {year} {2011})},\
  \Eprint {https://arxiv.org/abs/1104.1639} {arXiv:1104.1639 [astro-ph.SR]}
  \BibitemShut {NoStop}%
\bibitem [{\citenamefont {Escudero}\ \emph {et~al.}(2019)\citenamefont
  {Escudero}, \citenamefont {Hooper}, \citenamefont {Krnjaic},\ and\
  \citenamefont {Pierre}}]{Escudero:2019gzq}%
  \BibitemOpen
  \bibfield  {author} {\bibinfo {author} {\bibfnamefont {M.}~\bibnamefont
  {Escudero}}, \bibinfo {author} {\bibfnamefont {D.}~\bibnamefont {Hooper}},
  \bibinfo {author} {\bibfnamefont {G.}~\bibnamefont {Krnjaic}},\ and\ \bibinfo
  {author} {\bibfnamefont {M.}~\bibnamefont {Pierre}},\ }\href
  {https://doi.org/10.1007/JHEP03(2019)071} {\bibfield  {journal} {\bibinfo
  {journal} {JHEP}\ }\textbf {\bibinfo {volume} {03}},\ \bibinfo {pages}
  {071}},\ \Eprint {https://arxiv.org/abs/1901.02010} {arXiv:1901.02010
  [hep-ph]} \BibitemShut {NoStop}%
\bibitem [{\citenamefont {Croon}\ \emph {et~al.}(2021)\citenamefont {Croon},
  \citenamefont {Elor}, \citenamefont {Leane},\ and\ \citenamefont
  {McDermott}}]{Croon:2020lrf}%
  \BibitemOpen
  \bibfield  {author} {\bibinfo {author} {\bibfnamefont {D.}~\bibnamefont
  {Croon}}, \bibinfo {author} {\bibfnamefont {G.}~\bibnamefont {Elor}},
  \bibinfo {author} {\bibfnamefont {R.~K.}\ \bibnamefont {Leane}},\ and\
  \bibinfo {author} {\bibfnamefont {S.~D.}\ \bibnamefont {McDermott}},\ }\href
  {https://doi.org/10.1007/JHEP01(2021)107} {\bibfield  {journal} {\bibinfo
  {journal} {JHEP}\ }\textbf {\bibinfo {volume} {01}},\ \bibinfo {pages}
  {107}},\ \Eprint {https://arxiv.org/abs/2006.13942} {arXiv:2006.13942
  [hep-ph]} \BibitemShut {NoStop}%
\bibitem [{\citenamefont {Mishra}\ \emph {et~al.}(1991)\citenamefont {Mishra}
  \emph {et~al.}}]{CCFR:1991lpl}%
  \BibitemOpen
  \bibfield  {author} {\bibinfo {author} {\bibfnamefont {S.~R.}\ \bibnamefont
  {Mishra}} \emph {et~al.} (\bibinfo {collaboration} {CCFR}),\ }\href
  {https://doi.org/10.1103/PhysRevLett.66.3117} {\bibfield  {journal} {\bibinfo
   {journal} {Phys. Rev. Lett.}\ }\textbf {\bibinfo {volume} {66}},\ \bibinfo
  {pages} {3117} (\bibinfo {year} {1991})}\BibitemShut {NoStop}%
\bibitem [{\citenamefont {Altmannshofer}\ \emph {et~al.}(2014)\citenamefont
  {Altmannshofer}, \citenamefont {Gori}, \citenamefont {Pospelov},\ and\
  \citenamefont {Yavin}}]{Altmannshofer:2014pba}%
  \BibitemOpen
  \bibfield  {author} {\bibinfo {author} {\bibfnamefont {W.}~\bibnamefont
  {Altmannshofer}}, \bibinfo {author} {\bibfnamefont {S.}~\bibnamefont {Gori}},
  \bibinfo {author} {\bibfnamefont {M.}~\bibnamefont {Pospelov}},\ and\
  \bibinfo {author} {\bibfnamefont {I.}~\bibnamefont {Yavin}},\ }\href
  {https://doi.org/10.1103/PhysRevLett.113.091801} {\bibfield  {journal}
  {\bibinfo  {journal} {Phys. Rev. Lett.}\ }\textbf {\bibinfo {volume} {113}},\
  \bibinfo {pages} {091801} (\bibinfo {year} {2014})},\ \Eprint
  {https://arxiv.org/abs/1406.2332} {arXiv:1406.2332 [hep-ph]} \BibitemShut
  {NoStop}%
\bibitem [{\citenamefont {Cardoso}\ \emph {et~al.}(2018)\citenamefont
  {Cardoso}, \citenamefont {Dias}, \citenamefont {Hartnett}, \citenamefont
  {Middleton}, \citenamefont {Pani},\ and\ \citenamefont
  {Santos}}]{Cardoso:2018tly}%
  \BibitemOpen
  \bibfield  {author} {\bibinfo {author} {\bibfnamefont {V.}~\bibnamefont
  {Cardoso}}, \bibinfo {author} {\bibfnamefont {O.~J.~C.}\ \bibnamefont
  {Dias}}, \bibinfo {author} {\bibfnamefont {G.~S.}\ \bibnamefont {Hartnett}},
  \bibinfo {author} {\bibfnamefont {M.}~\bibnamefont {Middleton}}, \bibinfo
  {author} {\bibfnamefont {P.}~\bibnamefont {Pani}},\ and\ \bibinfo {author}
  {\bibfnamefont {J.~E.}\ \bibnamefont {Santos}},\ }\href
  {https://doi.org/10.1088/1475-7516/2018/03/043} {\bibfield  {journal}
  {\bibinfo  {journal} {JCAP}\ }\textbf {\bibinfo {volume} {03}},\ \bibinfo
  {pages} {043}},\ \Eprint {https://arxiv.org/abs/1801.01420} {arXiv:1801.01420
  [gr-qc]} \BibitemShut {NoStop}%
\bibitem [{\citenamefont {Stott}(2020)}]{Stott:2020gjj}%
  \BibitemOpen
  \bibfield  {author} {\bibinfo {author} {\bibfnamefont {M.~J.}\ \bibnamefont
  {Stott}},\ }\href@noop {} {\  (\bibinfo {year} {2020})},\ \Eprint
  {https://arxiv.org/abs/2009.07206} {arXiv:2009.07206 [hep-ph]} \BibitemShut
  {NoStop}%
\bibitem [{\citenamefont {Ghosh}\ and\ \citenamefont
  {Sachdeva}(2021)}]{Ghosh:2021zuf}%
  \BibitemOpen
  \bibfield  {author} {\bibinfo {author} {\bibfnamefont {D.}~\bibnamefont
  {Ghosh}}\ and\ \bibinfo {author} {\bibfnamefont {D.}~\bibnamefont
  {Sachdeva}},\ }\href {https://doi.org/10.1103/PhysRevD.103.095028} {\bibfield
   {journal} {\bibinfo  {journal} {Phys. Rev. D}\ }\textbf {\bibinfo {volume}
  {103}},\ \bibinfo {pages} {095028} (\bibinfo {year} {2021})},\ \Eprint
  {https://arxiv.org/abs/2102.08857} {arXiv:2102.08857 [astro-ph.HE]}
  \BibitemShut {NoStop}%
\bibitem [{\citenamefont {Czarnecki}\ and\ \citenamefont
  {Marciano}(2001)}]{Czarnecki:2001pv}%
  \BibitemOpen
  \bibfield  {author} {\bibinfo {author} {\bibfnamefont {A.}~\bibnamefont
  {Czarnecki}}\ and\ \bibinfo {author} {\bibfnamefont {W.~J.}\ \bibnamefont
  {Marciano}},\ }\href {https://doi.org/10.1103/PhysRevD.64.013014} {\bibfield
  {journal} {\bibinfo  {journal} {Phys. Rev. D}\ }\textbf {\bibinfo {volume}
  {64}},\ \bibinfo {pages} {013014} (\bibinfo {year} {2001})},\ \Eprint
  {https://arxiv.org/abs/hep-ph/0102122} {arXiv:hep-ph/0102122} \BibitemShut
  {NoStop}%
\bibitem [{\citenamefont {Baek}\ \emph {et~al.}(2001)\citenamefont {Baek},
  \citenamefont {Deshpande}, \citenamefont {He},\ and\ \citenamefont
  {Ko}}]{Baek:2001kca}%
  \BibitemOpen
  \bibfield  {author} {\bibinfo {author} {\bibfnamefont {S.}~\bibnamefont
  {Baek}}, \bibinfo {author} {\bibfnamefont {N.~G.}\ \bibnamefont {Deshpande}},
  \bibinfo {author} {\bibfnamefont {X.~G.}\ \bibnamefont {He}},\ and\ \bibinfo
  {author} {\bibfnamefont {P.}~\bibnamefont {Ko}},\ }\href
  {https://doi.org/10.1103/PhysRevD.64.055006} {\bibfield  {journal} {\bibinfo
  {journal} {Phys. Rev. D}\ }\textbf {\bibinfo {volume} {64}},\ \bibinfo
  {pages} {055006} (\bibinfo {year} {2001})},\ \Eprint
  {https://arxiv.org/abs/hep-ph/0104141} {arXiv:hep-ph/0104141} \BibitemShut
  {NoStop}%
\bibitem [{\citenamefont {Heeck}\ and\ \citenamefont
  {Thapa}(2022)}]{Heeck:2022znj}%
  \BibitemOpen
  \bibfield  {author} {\bibinfo {author} {\bibfnamefont {J.}~\bibnamefont
  {Heeck}}\ and\ \bibinfo {author} {\bibfnamefont {A.}~\bibnamefont {Thapa}},\
  }\href@noop {} {\  (\bibinfo {year} {2022})},\ \Eprint
  {https://arxiv.org/abs/2202.08854} {arXiv:2202.08854 [hep-ph]} \BibitemShut
  {NoStop}%
\bibitem [{\citenamefont {Schael}\ \emph {et~al.}(2013)\citenamefont {Schael}
  \emph {et~al.}}]{ALEPH:2013dgf}%
  \BibitemOpen
  \bibfield  {author} {\bibinfo {author} {\bibfnamefont {S.}~\bibnamefont
  {Schael}} \emph {et~al.} (\bibinfo {collaboration} {ALEPH, DELPHI, L3, OPAL,
  LEP Electroweak}),\ }\href {https://doi.org/10.1016/j.physrep.2013.07.004}
  {\bibfield  {journal} {\bibinfo  {journal} {Phys. Rept.}\ }\textbf {\bibinfo
  {volume} {532}},\ \bibinfo {pages} {119} (\bibinfo {year} {2013})},\ \Eprint
  {https://arxiv.org/abs/1302.3415} {arXiv:1302.3415 [hep-ex]} \BibitemShut
  {NoStop}%
\bibitem [{\citenamefont {Sirunyan}\ \emph {et~al.}(2019)\citenamefont
  {Sirunyan} \emph {et~al.}}]{CMS:2019ppl}%
  \BibitemOpen
  \bibfield  {author} {\bibinfo {author} {\bibfnamefont {A.~M.}\ \bibnamefont
  {Sirunyan}} \emph {et~al.} (\bibinfo {collaboration} {CMS}),\ }\href
  {https://doi.org/10.1007/JHEP12(2019)062} {\bibfield  {journal} {\bibinfo
  {journal} {JHEP}\ }\textbf {\bibinfo {volume} {12}},\ \bibinfo {pages}
  {062}},\ \Eprint {https://arxiv.org/abs/1907.08354} {arXiv:1907.08354
  [hep-ex]} \BibitemShut {NoStop}%
\bibitem [{\citenamefont {Schael}\ \emph {et~al.}(2006)\citenamefont {Schael}
  \emph {et~al.}}]{ALEPH:2005ab}%
  \BibitemOpen
  \bibfield  {author} {\bibinfo {author} {\bibfnamefont {S.}~\bibnamefont
  {Schael}} \emph {et~al.} (\bibinfo {collaboration} {ALEPH, DELPHI, L3, OPAL,
  SLD, LEP Electroweak Working Group, SLD Electroweak Group, SLD Heavy Flavour
  Group}),\ }\href {https://doi.org/10.1016/j.physrep.2005.12.006} {\bibfield
  {journal} {\bibinfo  {journal} {Phys. Rept.}\ }\textbf {\bibinfo {volume}
  {427}},\ \bibinfo {pages} {257} (\bibinfo {year} {2006})},\ \Eprint
  {https://arxiv.org/abs/hep-ex/0509008} {arXiv:hep-ex/0509008} \BibitemShut
  {NoStop}%
\bibitem [{\citenamefont {Smith}\ \emph {et~al.}(2000)\citenamefont {Smith},
  \citenamefont {Hoyle}, \citenamefont {Gundlach}, \citenamefont {Adelberger},
  \citenamefont {Heckel},\ and\ \citenamefont {Swanson}}]{Smith:1999cr}%
  \BibitemOpen
  \bibfield  {author} {\bibinfo {author} {\bibfnamefont {G.~L.}\ \bibnamefont
  {Smith}}, \bibinfo {author} {\bibfnamefont {C.~D.}\ \bibnamefont {Hoyle}},
  \bibinfo {author} {\bibfnamefont {J.~H.}\ \bibnamefont {Gundlach}}, \bibinfo
  {author} {\bibfnamefont {E.~G.}\ \bibnamefont {Adelberger}}, \bibinfo
  {author} {\bibfnamefont {B.~R.}\ \bibnamefont {Heckel}},\ and\ \bibinfo
  {author} {\bibfnamefont {H.~E.}\ \bibnamefont {Swanson}},\ }\href
  {https://doi.org/10.1103/PhysRevD.61.022001} {\bibfield  {journal} {\bibinfo
  {journal} {Phys. Rev. D}\ }\textbf {\bibinfo {volume} {61}},\ \bibinfo
  {pages} {022001} (\bibinfo {year} {2000})}\BibitemShut {NoStop}%
\bibitem [{\citenamefont {Fischbach}\ and\ \citenamefont
  {Talmadge}(1999)}]{Fischbach:1999bc}%
  \BibitemOpen
  \bibfield  {author} {\bibinfo {author} {\bibfnamefont {E.}~\bibnamefont
  {Fischbach}}\ and\ \bibinfo {author} {\bibfnamefont {C.~L.}\ \bibnamefont
  {Talmadge}},\ }\href@noop {} {\emph {\bibinfo {title} {{The search for
  nonNewtonian gravity}}}}\ (\bibinfo {year} {1999})\BibitemShut {NoStop}%
\bibitem [{\citenamefont {Adelberger}\ \emph {et~al.}(2009)\citenamefont
  {Adelberger}, \citenamefont {Gundlach}, \citenamefont {Heckel}, \citenamefont
  {Hoedl},\ and\ \citenamefont {Schlamminger}}]{Adelberger:2009zz}%
  \BibitemOpen
  \bibfield  {author} {\bibinfo {author} {\bibfnamefont {E.~G.}\ \bibnamefont
  {Adelberger}}, \bibinfo {author} {\bibfnamefont {J.~H.}\ \bibnamefont
  {Gundlach}}, \bibinfo {author} {\bibfnamefont {B.~R.}\ \bibnamefont
  {Heckel}}, \bibinfo {author} {\bibfnamefont {S.}~\bibnamefont {Hoedl}},\ and\
  \bibinfo {author} {\bibfnamefont {S.}~\bibnamefont {Schlamminger}},\ }\href
  {https://doi.org/10.1016/j.ppnp.2008.08.002} {\bibfield  {journal} {\bibinfo
  {journal} {Prog. Part. Nucl. Phys.}\ }\textbf {\bibinfo {volume} {62}},\
  \bibinfo {pages} {102} (\bibinfo {year} {2009})}\BibitemShut {NoStop}%
\bibitem [{\citenamefont {Wagner}\ \emph {et~al.}(2012)\citenamefont {Wagner},
  \citenamefont {Schlamminger}, \citenamefont {Gundlach},\ and\ \citenamefont
  {Adelberger}}]{Wagner:2012ui}%
  \BibitemOpen
  \bibfield  {author} {\bibinfo {author} {\bibfnamefont {T.~A.}\ \bibnamefont
  {Wagner}}, \bibinfo {author} {\bibfnamefont {S.}~\bibnamefont
  {Schlamminger}}, \bibinfo {author} {\bibfnamefont {J.~H.}\ \bibnamefont
  {Gundlach}},\ and\ \bibinfo {author} {\bibfnamefont {E.~G.}\ \bibnamefont
  {Adelberger}},\ }\href {https://doi.org/10.1088/0264-9381/29/18/184002}
  {\bibfield  {journal} {\bibinfo  {journal} {Class. Quant. Grav.}\ }\textbf
  {\bibinfo {volume} {29}},\ \bibinfo {pages} {184002} (\bibinfo {year}
  {2012})},\ \Eprint {https://arxiv.org/abs/1207.2442} {arXiv:1207.2442
  [gr-qc]} \BibitemShut {NoStop}%
\bibitem [{\citenamefont {Murata}\ and\ \citenamefont
  {Tanaka}(2015)}]{Murata:2014nra}%
  \BibitemOpen
  \bibfield  {author} {\bibinfo {author} {\bibfnamefont {J.}~\bibnamefont
  {Murata}}\ and\ \bibinfo {author} {\bibfnamefont {S.}~\bibnamefont
  {Tanaka}},\ }\href {https://doi.org/10.1088/0264-9381/32/3/033001} {\bibfield
   {journal} {\bibinfo  {journal} {Class. Quant. Grav.}\ }\textbf {\bibinfo
  {volume} {32}},\ \bibinfo {pages} {033001} (\bibinfo {year} {2015})},\
  \Eprint {https://arxiv.org/abs/1408.3588} {arXiv:1408.3588 [hep-ex]}
  \BibitemShut {NoStop}%
\bibitem [{\citenamefont {Fayet}(2018)}]{Fayet:2017pdp}%
  \BibitemOpen
  \bibfield  {author} {\bibinfo {author} {\bibfnamefont {P.}~\bibnamefont
  {Fayet}},\ }\href {https://doi.org/10.1103/PhysRevD.97.055039} {\bibfield
  {journal} {\bibinfo  {journal} {Phys. Rev. D}\ }\textbf {\bibinfo {volume}
  {97}},\ \bibinfo {pages} {055039} (\bibinfo {year} {2018})},\ \Eprint
  {https://arxiv.org/abs/1712.00856} {arXiv:1712.00856 [hep-ph]} \BibitemShut
  {NoStop}%
\bibitem [{\citenamefont {Shin}\ and\ \citenamefont
  {Yun}(2022)}]{Shin:2021bvz}%
  \BibitemOpen
  \bibfield  {author} {\bibinfo {author} {\bibfnamefont {C.~S.}\ \bibnamefont
  {Shin}}\ and\ \bibinfo {author} {\bibfnamefont {S.}~\bibnamefont {Yun}},\
  }\href {https://doi.org/10.1007/JHEP02(2022)133} {\bibfield  {journal}
  {\bibinfo  {journal} {JHEP}\ }\textbf {\bibinfo {volume} {02}},\ \bibinfo
  {pages} {133}},\ \Eprint {https://arxiv.org/abs/2110.03362} {arXiv:2110.03362
  [hep-ph]} \BibitemShut {NoStop}%
\bibitem [{\citenamefont {An}\ \emph {et~al.}(2013)\citenamefont {An},
  \citenamefont {Pospelov},\ and\ \citenamefont {Pradler}}]{An:2013yfc}%
  \BibitemOpen
  \bibfield  {author} {\bibinfo {author} {\bibfnamefont {H.}~\bibnamefont
  {An}}, \bibinfo {author} {\bibfnamefont {M.}~\bibnamefont {Pospelov}},\ and\
  \bibinfo {author} {\bibfnamefont {J.}~\bibnamefont {Pradler}},\ }\href
  {https://doi.org/10.1016/j.physletb.2013.07.008} {\bibfield  {journal}
  {\bibinfo  {journal} {Phys. Lett. B}\ }\textbf {\bibinfo {volume} {725}},\
  \bibinfo {pages} {190} (\bibinfo {year} {2013})},\ \Eprint
  {https://arxiv.org/abs/1302.3884} {arXiv:1302.3884 [hep-ph]} \BibitemShut
  {NoStop}%
\bibitem [{\citenamefont {Redondo}\ and\ \citenamefont
  {Raffelt}(2013)}]{Redondo:2013lna}%
  \BibitemOpen
  \bibfield  {author} {\bibinfo {author} {\bibfnamefont {J.}~\bibnamefont
  {Redondo}}\ and\ \bibinfo {author} {\bibfnamefont {G.}~\bibnamefont
  {Raffelt}},\ }\href {https://doi.org/10.1088/1475-7516/2013/08/034}
  {\bibfield  {journal} {\bibinfo  {journal} {JCAP}\ }\textbf {\bibinfo
  {volume} {08}},\ \bibinfo {pages} {034}},\ \Eprint
  {https://arxiv.org/abs/1305.2920} {arXiv:1305.2920 [hep-ph]} \BibitemShut
  {NoStop}%
\bibitem [{\citenamefont {Raffelt}(1996)}]{Raffelt:1996wa}%
  \BibitemOpen
  \bibfield  {author} {\bibinfo {author} {\bibfnamefont {G.~G.}\ \bibnamefont
  {Raffelt}},\ }\href@noop {} {\emph {\bibinfo {title} {{Stars as laboratories
  for fundamental physics}: {The astrophysics of neutrinos, axions, and other
  weakly interacting particles}}}}\ (\bibinfo {year} {1996})\BibitemShut
  {NoStop}%
\bibitem [{\citenamefont {Brust}\ \emph {et~al.}(2013)\citenamefont {Brust},
  \citenamefont {Kaplan},\ and\ \citenamefont {Walters}}]{Brust:2013ova}%
  \BibitemOpen
  \bibfield  {author} {\bibinfo {author} {\bibfnamefont {C.}~\bibnamefont
  {Brust}}, \bibinfo {author} {\bibfnamefont {D.~E.}\ \bibnamefont {Kaplan}},\
  and\ \bibinfo {author} {\bibfnamefont {M.~T.}\ \bibnamefont {Walters}},\
  }\href {https://doi.org/10.1007/JHEP12(2013)058} {\bibfield  {journal}
  {\bibinfo  {journal} {JHEP}\ }\textbf {\bibinfo {volume} {12}},\ \bibinfo
  {pages} {058}},\ \Eprint {https://arxiv.org/abs/1303.5379} {arXiv:1303.5379
  [hep-ph]} \BibitemShut {NoStop}%
\bibitem [{\citenamefont {Bollig}\ \emph {et~al.}(2017)\citenamefont {Bollig},
  \citenamefont {Janka}, \citenamefont {Lohs}, \citenamefont {Martinez-Pinedo},
  \citenamefont {Horowitz},\ and\ \citenamefont {Melson}}]{Bollig:2017lki}%
  \BibitemOpen
  \bibfield  {author} {\bibinfo {author} {\bibfnamefont {R.}~\bibnamefont
  {Bollig}}, \bibinfo {author} {\bibfnamefont {H.~T.}\ \bibnamefont {Janka}},
  \bibinfo {author} {\bibfnamefont {A.}~\bibnamefont {Lohs}}, \bibinfo {author}
  {\bibfnamefont {G.}~\bibnamefont {Martinez-Pinedo}}, \bibinfo {author}
  {\bibfnamefont {C.~J.}\ \bibnamefont {Horowitz}},\ and\ \bibinfo {author}
  {\bibfnamefont {T.}~\bibnamefont {Melson}},\ }\href
  {https://doi.org/10.1103/PhysRevLett.119.242702} {\bibfield  {journal}
  {\bibinfo  {journal} {Phys. Rev. Lett.}\ }\textbf {\bibinfo {volume} {119}},\
  \bibinfo {pages} {242702} (\bibinfo {year} {2017})},\ \Eprint
  {https://arxiv.org/abs/1706.04630} {arXiv:1706.04630 [astro-ph.HE]}
  \BibitemShut {NoStop}%
\bibitem [{\citenamefont {Kumar~Poddar}\ \emph {et~al.}(2019)\citenamefont
  {Kumar~Poddar}, \citenamefont {Mohanty},\ and\ \citenamefont
  {Jana}}]{KumarPoddar:2019ceq}%
  \BibitemOpen
  \bibfield  {author} {\bibinfo {author} {\bibfnamefont {T.}~\bibnamefont
  {Kumar~Poddar}}, \bibinfo {author} {\bibfnamefont {S.}~\bibnamefont
  {Mohanty}},\ and\ \bibinfo {author} {\bibfnamefont {S.}~\bibnamefont
  {Jana}},\ }\href {https://doi.org/10.1103/PhysRevD.100.123023} {\bibfield
  {journal} {\bibinfo  {journal} {Phys. Rev. D}\ }\textbf {\bibinfo {volume}
  {100}},\ \bibinfo {pages} {123023} (\bibinfo {year} {2019})},\ \Eprint
  {https://arxiv.org/abs/1908.09732} {arXiv:1908.09732 [hep-ph]} \BibitemShut
  {NoStop}%
\bibitem [{\citenamefont {Hardy}\ and\ \citenamefont
  {Lasenby}(2017)}]{Hardy:2016kme}%
  \BibitemOpen
  \bibfield  {author} {\bibinfo {author} {\bibfnamefont {E.}~\bibnamefont
  {Hardy}}\ and\ \bibinfo {author} {\bibfnamefont {R.}~\bibnamefont
  {Lasenby}},\ }\href {https://doi.org/10.1007/JHEP02(2017)033} {\bibfield
  {journal} {\bibinfo  {journal} {JHEP}\ }\textbf {\bibinfo {volume} {02}},\
  \bibinfo {pages} {033}},\ \Eprint {https://arxiv.org/abs/1611.05852}
  {arXiv:1611.05852 [hep-ph]} \BibitemShut {NoStop}%
\bibitem [{\citenamefont {Hong}\ \emph {et~al.}(2021)\citenamefont {Hong},
  \citenamefont {Shin},\ and\ \citenamefont {Yun}}]{Hong:2020bxo}%
  \BibitemOpen
  \bibfield  {author} {\bibinfo {author} {\bibfnamefont {D.~K.}\ \bibnamefont
  {Hong}}, \bibinfo {author} {\bibfnamefont {C.~S.}\ \bibnamefont {Shin}},\
  and\ \bibinfo {author} {\bibfnamefont {S.}~\bibnamefont {Yun}},\ }\href
  {https://doi.org/10.1103/PhysRevD.103.123031} {\bibfield  {journal} {\bibinfo
   {journal} {Phys. Rev. D}\ }\textbf {\bibinfo {volume} {103}},\ \bibinfo
  {pages} {123031} (\bibinfo {year} {2021})},\ \Eprint
  {https://arxiv.org/abs/2012.05427} {arXiv:2012.05427 [hep-ph]} \BibitemShut
  {NoStop}%
\bibitem [{\citenamefont {Dubovsky}\ and\ \citenamefont
  {Hern\'andez-Chifflet}(2015)}]{Dubovsky:2015cca}%
  \BibitemOpen
  \bibfield  {author} {\bibinfo {author} {\bibfnamefont {S.}~\bibnamefont
  {Dubovsky}}\ and\ \bibinfo {author} {\bibfnamefont {G.}~\bibnamefont
  {Hern\'andez-Chifflet}},\ }\href
  {https://doi.org/10.1088/1475-7516/2015/12/054} {\bibfield  {journal}
  {\bibinfo  {journal} {JCAP}\ }\textbf {\bibinfo {volume} {12}},\ \bibinfo
  {pages} {054}},\ \Eprint {https://arxiv.org/abs/1509.00039} {arXiv:1509.00039
  [hep-ph]} \BibitemShut {NoStop}%
\bibitem [{\citenamefont {Wadekar}\ and\ \citenamefont
  {Farrar}(2021)}]{Wadekar:2019mpc}%
  \BibitemOpen
  \bibfield  {author} {\bibinfo {author} {\bibfnamefont {D.}~\bibnamefont
  {Wadekar}}\ and\ \bibinfo {author} {\bibfnamefont {G.~R.}\ \bibnamefont
  {Farrar}},\ }\href {https://doi.org/10.1103/PhysRevD.103.123028} {\bibfield
  {journal} {\bibinfo  {journal} {Phys. Rev. D}\ }\textbf {\bibinfo {volume}
  {103}},\ \bibinfo {pages} {123028} (\bibinfo {year} {2021})},\ \Eprint
  {https://arxiv.org/abs/1903.12190} {arXiv:1903.12190 [hep-ph]} \BibitemShut
  {NoStop}%
\bibitem [{\citenamefont {Bhoonah}\ \emph {et~al.}(2020)\citenamefont
  {Bhoonah}, \citenamefont {Bramante},\ and\ \citenamefont
  {Song}}]{Bhoonah:2019eyo}%
  \BibitemOpen
  \bibfield  {author} {\bibinfo {author} {\bibfnamefont {A.}~\bibnamefont
  {Bhoonah}}, \bibinfo {author} {\bibfnamefont {J.}~\bibnamefont {Bramante}},\
  and\ \bibinfo {author} {\bibfnamefont {N.}~\bibnamefont {Song}},\ }\href
  {https://doi.org/10.1103/PhysRevD.101.055040} {\bibfield  {journal} {\bibinfo
   {journal} {Phys. Rev. D}\ }\textbf {\bibinfo {volume} {101}},\ \bibinfo
  {pages} {055040} (\bibinfo {year} {2020})},\ \Eprint
  {https://arxiv.org/abs/1909.07387} {arXiv:1909.07387 [hep-ph]} \BibitemShut
  {NoStop}%
\bibitem [{\citenamefont {McDermott}\ and\ \citenamefont
  {Witte}(2020)}]{McDermott:2019lch}%
  \BibitemOpen
  \bibfield  {author} {\bibinfo {author} {\bibfnamefont {S.~D.}\ \bibnamefont
  {McDermott}}\ and\ \bibinfo {author} {\bibfnamefont {S.~J.}\ \bibnamefont
  {Witte}},\ }\href {https://doi.org/10.1103/PhysRevD.101.063030} {\bibfield
  {journal} {\bibinfo  {journal} {Phys. Rev. D}\ }\textbf {\bibinfo {volume}
  {101}},\ \bibinfo {pages} {063030} (\bibinfo {year} {2020})},\ \Eprint
  {https://arxiv.org/abs/1911.05086} {arXiv:1911.05086 [hep-ph]} \BibitemShut
  {NoStop}%
\bibitem [{\citenamefont {Caputo}\ \emph
  {et~al.}(2020{\natexlab{a}})\citenamefont {Caputo}, \citenamefont {Liu},
  \citenamefont {Mishra-Sharma},\ and\ \citenamefont
  {Ruderman}}]{Caputo:2020bdy}%
  \BibitemOpen
  \bibfield  {author} {\bibinfo {author} {\bibfnamefont {A.}~\bibnamefont
  {Caputo}}, \bibinfo {author} {\bibfnamefont {H.}~\bibnamefont {Liu}},
  \bibinfo {author} {\bibfnamefont {S.}~\bibnamefont {Mishra-Sharma}},\ and\
  \bibinfo {author} {\bibfnamefont {J.~T.}\ \bibnamefont {Ruderman}},\ }\href
  {https://doi.org/10.1103/PhysRevLett.125.221303} {\bibfield  {journal}
  {\bibinfo  {journal} {Phys. Rev. Lett.}\ }\textbf {\bibinfo {volume} {125}},\
  \bibinfo {pages} {221303} (\bibinfo {year} {2020}{\natexlab{a}})},\ \Eprint
  {https://arxiv.org/abs/2002.05165} {arXiv:2002.05165 [astro-ph.CO]}
  \BibitemShut {NoStop}%
\bibitem [{\citenamefont {Witte}\ \emph {et~al.}(2020)\citenamefont {Witte},
  \citenamefont {Rosauro-Alcaraz}, \citenamefont {McDermott},\ and\
  \citenamefont {Poulin}}]{Witte:2020rvb}%
  \BibitemOpen
  \bibfield  {author} {\bibinfo {author} {\bibfnamefont {S.~J.}\ \bibnamefont
  {Witte}}, \bibinfo {author} {\bibfnamefont {S.}~\bibnamefont
  {Rosauro-Alcaraz}}, \bibinfo {author} {\bibfnamefont {S.~D.}\ \bibnamefont
  {McDermott}},\ and\ \bibinfo {author} {\bibfnamefont {V.}~\bibnamefont
  {Poulin}},\ }\href {https://doi.org/10.1007/JHEP06(2020)132} {\bibfield
  {journal} {\bibinfo  {journal} {JHEP}\ }\textbf {\bibinfo {volume} {06}},\
  \bibinfo {pages} {132}},\ \Eprint {https://arxiv.org/abs/2003.13698}
  {arXiv:2003.13698 [astro-ph.CO]} \BibitemShut {NoStop}%
\bibitem [{\citenamefont {Caputo}\ \emph
  {et~al.}(2020{\natexlab{b}})\citenamefont {Caputo}, \citenamefont {Liu},
  \citenamefont {Mishra-Sharma},\ and\ \citenamefont
  {Ruderman}}]{Caputo:2020rnx}%
  \BibitemOpen
  \bibfield  {author} {\bibinfo {author} {\bibfnamefont {A.}~\bibnamefont
  {Caputo}}, \bibinfo {author} {\bibfnamefont {H.}~\bibnamefont {Liu}},
  \bibinfo {author} {\bibfnamefont {S.}~\bibnamefont {Mishra-Sharma}},\ and\
  \bibinfo {author} {\bibfnamefont {J.~T.}\ \bibnamefont {Ruderman}},\ }\href
  {https://doi.org/10.1103/PhysRevD.102.103533} {\bibfield  {journal} {\bibinfo
   {journal} {Phys. Rev. D}\ }\textbf {\bibinfo {volume} {102}},\ \bibinfo
  {pages} {103533} (\bibinfo {year} {2020}{\natexlab{b}})},\ \Eprint
  {https://arxiv.org/abs/2004.06733} {arXiv:2004.06733 [astro-ph.CO]}
  \BibitemShut {NoStop}%
\bibitem [{\citenamefont {Chang}\ \emph {et~al.}(2017)\citenamefont {Chang},
  \citenamefont {Essig},\ and\ \citenamefont {McDermott}}]{Chang:2016ntp}%
  \BibitemOpen
  \bibfield  {author} {\bibinfo {author} {\bibfnamefont {J.~H.}\ \bibnamefont
  {Chang}}, \bibinfo {author} {\bibfnamefont {R.}~\bibnamefont {Essig}},\ and\
  \bibinfo {author} {\bibfnamefont {S.~D.}\ \bibnamefont {McDermott}},\ }\href
  {https://doi.org/10.1007/JHEP01(2017)107} {\bibfield  {journal} {\bibinfo
  {journal} {JHEP}\ }\textbf {\bibinfo {volume} {01}},\ \bibinfo {pages}
  {107}},\ \Eprint {https://arxiv.org/abs/1611.03864} {arXiv:1611.03864
  [hep-ph]} \BibitemShut {NoStop}%
\bibitem [{\citenamefont {Povey}\ \emph {et~al.}(2010)\citenamefont {Povey},
  \citenamefont {Hartnett},\ and\ \citenamefont {Tobar}}]{Povey:2010hs}%
  \BibitemOpen
  \bibfield  {author} {\bibinfo {author} {\bibfnamefont {R.}~\bibnamefont
  {Povey}}, \bibinfo {author} {\bibfnamefont {J.}~\bibnamefont {Hartnett}},\
  and\ \bibinfo {author} {\bibfnamefont {M.}~\bibnamefont {Tobar}},\ }\href
  {https://doi.org/10.1103/PhysRevD.82.052003} {\bibfield  {journal} {\bibinfo
  {journal} {Phys. Rev. D}\ }\textbf {\bibinfo {volume} {82}},\ \bibinfo
  {pages} {052003} (\bibinfo {year} {2010})},\ \Eprint
  {https://arxiv.org/abs/1003.0964} {arXiv:1003.0964 [hep-ex]} \BibitemShut
  {NoStop}%
\bibitem [{\citenamefont {Ehret}\ \emph {et~al.}(2010)\citenamefont {Ehret}
  \emph {et~al.}}]{Ehret:2010mh}%
  \BibitemOpen
  \bibfield  {author} {\bibinfo {author} {\bibfnamefont {K.}~\bibnamefont
  {Ehret}} \emph {et~al.},\ }\href
  {https://doi.org/10.1016/j.physletb.2010.04.066} {\bibfield  {journal}
  {\bibinfo  {journal} {Phys. Lett. B}\ }\textbf {\bibinfo {volume} {689}},\
  \bibinfo {pages} {149} (\bibinfo {year} {2010})},\ \Eprint
  {https://arxiv.org/abs/1004.1313} {arXiv:1004.1313 [hep-ex]} \BibitemShut
  {NoStop}%
\bibitem [{\citenamefont {Wagner}\ \emph {et~al.}(2010)\citenamefont {Wagner}
  \emph {et~al.}}]{ADMX:2010ubl}%
  \BibitemOpen
  \bibfield  {author} {\bibinfo {author} {\bibfnamefont {A.}~\bibnamefont
  {Wagner}} \emph {et~al.} (\bibinfo {collaboration} {ADMX}),\ }\href
  {https://doi.org/10.1103/PhysRevLett.105.171801} {\bibfield  {journal}
  {\bibinfo  {journal} {Phys. Rev. Lett.}\ }\textbf {\bibinfo {volume} {105}},\
  \bibinfo {pages} {171801} (\bibinfo {year} {2010})},\ \Eprint
  {https://arxiv.org/abs/1007.3766} {arXiv:1007.3766 [hep-ex]} \BibitemShut
  {NoStop}%
\bibitem [{\citenamefont {Inada}\ \emph {et~al.}(2013)\citenamefont {Inada},
  \citenamefont {Namba}, \citenamefont {Asai}, \citenamefont {Kobayashi},
  \citenamefont {Tanaka}, \citenamefont {Tamasaku}, \citenamefont {Sawada},\
  and\ \citenamefont {Ishikawa}}]{Inada:2013tx}%
  \BibitemOpen
  \bibfield  {author} {\bibinfo {author} {\bibfnamefont {T.}~\bibnamefont
  {Inada}}, \bibinfo {author} {\bibfnamefont {T.}~\bibnamefont {Namba}},
  \bibinfo {author} {\bibfnamefont {S.}~\bibnamefont {Asai}}, \bibinfo {author}
  {\bibfnamefont {T.}~\bibnamefont {Kobayashi}}, \bibinfo {author}
  {\bibfnamefont {Y.}~\bibnamefont {Tanaka}}, \bibinfo {author} {\bibfnamefont
  {K.}~\bibnamefont {Tamasaku}}, \bibinfo {author} {\bibfnamefont
  {K.}~\bibnamefont {Sawada}},\ and\ \bibinfo {author} {\bibfnamefont
  {T.}~\bibnamefont {Ishikawa}},\ }\href
  {https://doi.org/10.1016/j.physletb.2013.04.033} {\bibfield  {journal}
  {\bibinfo  {journal} {Phys. Lett. B}\ }\textbf {\bibinfo {volume} {722}},\
  \bibinfo {pages} {301} (\bibinfo {year} {2013})},\ \Eprint
  {https://arxiv.org/abs/1301.6557} {arXiv:1301.6557 [physics.ins-det]}
  \BibitemShut {NoStop}%
\bibitem [{\citenamefont {Betz}\ \emph {et~al.}(2013)\citenamefont {Betz},
  \citenamefont {Caspers}, \citenamefont {Gasior}, \citenamefont {Thumm},\ and\
  \citenamefont {Rieger}}]{Betz:2013dza}%
  \BibitemOpen
  \bibfield  {author} {\bibinfo {author} {\bibfnamefont {M.}~\bibnamefont
  {Betz}}, \bibinfo {author} {\bibfnamefont {F.}~\bibnamefont {Caspers}},
  \bibinfo {author} {\bibfnamefont {M.}~\bibnamefont {Gasior}}, \bibinfo
  {author} {\bibfnamefont {M.}~\bibnamefont {Thumm}},\ and\ \bibinfo {author}
  {\bibfnamefont {S.~W.}\ \bibnamefont {Rieger}},\ }\href
  {https://doi.org/10.1103/PhysRevD.88.075014} {\bibfield  {journal} {\bibinfo
  {journal} {Phys. Rev. D}\ }\textbf {\bibinfo {volume} {88}},\ \bibinfo
  {pages} {075014} (\bibinfo {year} {2013})},\ \Eprint
  {https://arxiv.org/abs/1310.8098} {arXiv:1310.8098 [physics.ins-det]}
  \BibitemShut {NoStop}%
\bibitem [{\citenamefont {Parker}\ \emph {et~al.}(2013)\citenamefont {Parker},
  \citenamefont {Hartnett}, \citenamefont {Povey},\ and\ \citenamefont
  {Tobar}}]{Parker:2013fxa}%
  \BibitemOpen
  \bibfield  {author} {\bibinfo {author} {\bibfnamefont {S.~R.}\ \bibnamefont
  {Parker}}, \bibinfo {author} {\bibfnamefont {J.~G.}\ \bibnamefont
  {Hartnett}}, \bibinfo {author} {\bibfnamefont {R.~G.}\ \bibnamefont
  {Povey}},\ and\ \bibinfo {author} {\bibfnamefont {M.~E.}\ \bibnamefont
  {Tobar}},\ }\href {https://doi.org/10.1103/PhysRevD.88.112004} {\bibfield
  {journal} {\bibinfo  {journal} {Phys. Rev. D}\ }\textbf {\bibinfo {volume}
  {88}},\ \bibinfo {pages} {112004} (\bibinfo {year} {2013})},\ \Eprint
  {https://arxiv.org/abs/1410.5244} {arXiv:1410.5244 [hep-ex]} \BibitemShut
  {NoStop}%
\bibitem [{\citenamefont {Ahonen}\ \emph {et~al.}(1996)\citenamefont {Ahonen},
  \citenamefont {Enqvist},\ and\ \citenamefont {Raffelt}}]{Ahonen:1995ky}%
  \BibitemOpen
  \bibfield  {author} {\bibinfo {author} {\bibfnamefont {J.}~\bibnamefont
  {Ahonen}}, \bibinfo {author} {\bibfnamefont {K.}~\bibnamefont {Enqvist}},\
  and\ \bibinfo {author} {\bibfnamefont {G.}~\bibnamefont {Raffelt}},\ }\href
  {https://doi.org/10.1016/0370-2693(95)01393-8} {\bibfield  {journal}
  {\bibinfo  {journal} {Phys. Lett. B}\ }\textbf {\bibinfo {volume} {366}},\
  \bibinfo {pages} {224} (\bibinfo {year} {1996})},\ \Eprint
  {https://arxiv.org/abs/hep-ph/9510211} {arXiv:hep-ph/9510211} \BibitemShut
  {NoStop}%
\bibitem [{\citenamefont {Mirizzi}\ \emph {et~al.}(2009)\citenamefont
  {Mirizzi}, \citenamefont {Redondo},\ and\ \citenamefont
  {Sigl}}]{Mirizzi:2009iz}%
  \BibitemOpen
  \bibfield  {author} {\bibinfo {author} {\bibfnamefont {A.}~\bibnamefont
  {Mirizzi}}, \bibinfo {author} {\bibfnamefont {J.}~\bibnamefont {Redondo}},\
  and\ \bibinfo {author} {\bibfnamefont {G.}~\bibnamefont {Sigl}},\ }\href
  {https://doi.org/10.1088/1475-7516/2009/03/026} {\bibfield  {journal}
  {\bibinfo  {journal} {JCAP}\ }\textbf {\bibinfo {volume} {03}},\ \bibinfo
  {pages} {026}},\ \Eprint {https://arxiv.org/abs/0901.0014} {arXiv:0901.0014
  [hep-ph]} \BibitemShut {NoStop}%
\bibitem [{\citenamefont {Amin}\ \emph {et~al.}(2022)\citenamefont {Amin},
  \citenamefont {Jain}, \citenamefont {Karur},\ and\ \citenamefont
  {Mocz}}]{Amin:2022pzv}%
  \BibitemOpen
  \bibfield  {author} {\bibinfo {author} {\bibfnamefont {M.~A.}\ \bibnamefont
  {Amin}}, \bibinfo {author} {\bibfnamefont {M.}~\bibnamefont {Jain}}, \bibinfo
  {author} {\bibfnamefont {R.}~\bibnamefont {Karur}},\ and\ \bibinfo {author}
  {\bibfnamefont {P.}~\bibnamefont {Mocz}},\ }\href@noop {} {\  (\bibinfo
  {year} {2022})},\ \Eprint {https://arxiv.org/abs/2203.11935}
  {arXiv:2203.11935 [astro-ph.CO]} \BibitemShut {NoStop}%
\end{thebibliography}%

\end{document}